\documentclass[a4paper,fleqn,usenatbib]{mnras}

\usepackage[T1]{fontenc}
\usepackage{ae,aecompl}
\usepackage{ulem}

\usepackage{graphicx}	% Including figure files
\usepackage{amsmath}	% Advanced maths commands
\usepackage{amssymb}	% Extra maths symbols
\usepackage{pdflscape}
\usepackage{natbib}
\title[Chemical abundances of primary stars in the Sirius-like binary systems]{Chemical abundances of primary stars in the Sirius-like binary systems}

\author[X. M. Kong et al.]{
 X. M. Kong,$^{1,2,3}$\thanks{Contact e-mail: \href{mailto:xmkong@nao.cas.cn}{xmkong@nao.cas.cn}}
 G. Zhao,$^{1,3}$\thanks{Contact e-mail: \href{mailto:gzhao@nao.cas.cn}{gzhao@nao.cas.cn}}
 J. K. Zhao,$^{1}$
 J. R. Shi, $^{1,3}$
 Y. Bharat Kumar,$^{1}$
 \newauthor L. Wang, $^{1}$
J. B. Zhang, $^{1}$
 Y. Wang, $^{1}$
  and Y. T. Zhou $^{1,3}$
 \\ \\
 $^{1}$ Key Laboratory of Optical Astronomy, National Astronomical Observatories, Chinese Academy of Sciences, Beijing 100012, China \\
 $^{2}$ School of Mechanical, Electrical and Information Engineering, Shandong University at Weihai, Weihai 264209, China \\
 $^{3}$ School of Astronomy and Space Science, University of Chinese Academy of Sciences, Beijing 100049, China
 }
 \date{Accepted XXX. Received YYY; in original form ZZZ}

\pubyear{2017}
\begin{document}
\label{firstpage}
\pagerange{\pageref{firstpage}--\pageref{lastpage}}
\maketitle

% Abstract of the paper
\begin{abstract}
Study of primary stars lying in Sirius-like systems with various masses of WD companions and orbital separations is one of the key aspects to understand the origin and nature of Barium (Ba) stars. In this paper, based on high resolution and high S/N spectra, we present systematic analysis of photospheric abundances for 18 FGK primary stars of Sirius-like systems including six giants and 12 dwarfs. Atmospheric parameters, stellar masses, and abundances of 24 elements (C, Na, Mg, Al, Si, S, K, Ca, Sc, Ti, V, Cr, Mn, Fe, Co, Ni, Cu, Sr, Y, Zr, Ba, La, Ce and Nd) are determined homogeneously. The abundance patterns in these sample stars show that most of the elements in our sample follow the behavior of field stars with similar metallicity.
As expected, s-process elements in four known Ba giants show overabundance. A weak correlation was found between anomalies of s-process elemental abundance and orbital separation, suggesting the orbital separation of the binaries could not be the main constraint to differentiate strong Ba stars from mild Ba stars. Our study shows that the large mass ($>$0.51 M$_{\odot}$) of a WD companion in a binary system is not a sufficient condition to form a Ba star, even if the separation between the two components is small. Although not sufficient it seems to be a necessary condition since Ba stars with lower mass WDs in the observed sample were not found.
Our results support that [s/Fe] and [hs/ls] ratios of Ba stars are anti-correlated with the metallicity. However, the different levels of s-process overabundance among Ba stars may not to be dominated mainly by the metallicity.

\end{abstract}

% Select between one and six entries from the list of approved keywords.
% Don't make up new ones.
\begin{keywords}
stars: fundamental parameters -- stars: abundances -- stars: chemically peculiar-- (stars:) binaries: general -- (stars:) white dwarfs
\end{keywords}

%%%%%%%%%%%%%%%%%%%%%%%%%%%%%%%%%%%%%%%%%%%%%%%%%%

%%%%%%%%%%%%%%%%% BODY OF PAPER %%%%%%%%%%%%%%%%%%

\section{Introduction}

Sirius-like systems (SLSs), taking Sirius as the prototype, refer to white dwarfs (WDs) in binary or multiple star systems that contain at least one less luminous companion of spectral type K or earlier \citep{Holberg2013}. These systems play an important role in astrophysics, and are crucial for the studies of stellar and galactic evolution: 1) They are often used to investigate the initial-final mass relation (IFMR) among WDs \citep{Zhao2012}; 2) They are important to study the mass and mass-ratio properties of main-sequence stars in binary systems \citep{Ferrario2012}; 3) Their study provides constraints on the physics of common-envelope evolution \citep{Zorotovic2010}; 4) They are useful to study the progenitor model for type Ia supernovae \citep{2010W}; 5) They can be used to investigate the origin and nature of Ba stars.

Barium~II (Ba) stars are G-K type giants (e.g. \citealt{Bidelman1951}) and dwarfs (e.g. \citealt{Tomkin1989}) that are spectroscopically characterized by overabundance of Ba and other s-process elements. The s-process nucleosynthesis is expected in the interiors of thermally pulsating Asymptotic Giant Branch (TP-AGB) stars followed by deep dredge-up phenomena, named third dredge-up, which brings the processed materials to the surface. However, the luminosity and temperature estimates for the Ba stars are not expected for them to evolve far enough to be able to produce the s-process elements in their interiors, thus they cannot be self-enriched \citep{Smiljanic2007}. In the early eighties, \cite{McClure1980} discovered that all Ba giants are likely members of binary systems through radial velocity monitoring. Further studies \citep{McClure1983, Boffin1988, Jorissen1998} showed that the companions could be white dwarfs. So, it is generally believed that Ba stars belong to binary systems and the chemical peculiarities observed in them are due to mass transfer from their AGB companions (now WD companions) during TP-AGB phase. Therefore, a binary system with a WD companion appears to be a necessary condition for producing a Ba star. Consequently, a series of critical questions aroused the scholars' enormous interest. One is that whether the existence of a WD companion in a binary system is sufficient to produce a Ba star. The identification of some normal giants with WD companions gave a negative answer to this question \citep{Jorissen1992, Zacs1997, Merle2014}, suggested to look for other important factors to understand the chemical peculiarities observed in Ba stars. \cite{Boffin1988} suggested that the degree of chemical contamination in Ba stars depends on the mass of the WD companion and the orbital separation in the binaries. \cite{Hurley2000} predicted that the minimum WD mass should be at least 0.51 M$_\odot$ so as to its progenitor can evolve to the AGB phase. Merle et al. (2016, hereafter MER16) studied 11 binary systems involving WD companions of various masses, wherein three primary stars with WD companions of masses lower than about 0.5 M$_\odot$ indeed did not show s-process overabundances, and six giants among eight sample stars with WD companions of mass above 0.5 M$_{\odot}$ showed clear s-process enrichment. Such study supports the prediction of \cite{Hurley2000},
 %whose WD mass qualifies them for being polluted by s-process material, but are not}
 however, the constraints are provided based on few observed data points and hence required larger sample to draw any firm conclusion.

The level of chemical enrichment in Ba stars and connection with their orbital period is investigated extensively and showed that the degree of Ba pollution is strongly correlated with the orbital period and mild Ba stars should have wider orbital separations than classical Ba stars \citep{Zacs1994, Han1995, Antipova2004, Yang2016}. On the other hand, \cite{Jorissen1998} suggested that the difference between mild and strong Ba stars may not be dominated by the orbital separation, and metallicity could play important role. This argument supports the study of \cite{Kovacs1985} who suggested that increase in Ba enrichment in stars follows with iron deficiency. \cite{Smiljanic2007} claimed that the neutron exposure during the s-process operation may play an important role. Other parameters, such as eccentricity, mass-loss mechanism (wind accretion or Roche lobe overflow), the efficiency of thermal pulses and dilution factors, also play role in observed chemical peculiarities of Ba stars \citep{, 2017Swaelmen, Busso2001, Allen2006, Castro2016}.

In general, the origin of s-process elements enrichment in Ba stars is still not well understood and lacks a definite conclusion. Detailed abundance analysis of large sample of binaries with a variety of orbital periods and masses of WD companions may give clues to their origin. In this paper, we analyzed the elemental abundances of 18 FGK primary stars in SLSs including four known Ba giants, two giants and 12 dwarfs. Together with three Ba dwarfs lying in SLSs which have been reported in \cite{2017Kong} (hereafter paper I), we compared the abundance patterns from this study with the general trends observed among normal disk stars from literature, and discussed the role of WD companion masses,  orbital period and the metallicity in the levels of s-process overabundance in Ba stars.
Our paper is organized as follows, Section 2 presents the sample selection. Observations and process of data reduction are described in Section 3. Analysis and results are presented in Section 4. The results are discussed in Section 5, and conclusions are drawn in Section 6.

\section{Sample selection}
We have selected 21 primary stars ( including BD+68$^\circ$1027, RE~J0702+129 and BD+80$^\circ$670 from Paper I) of FGK type in SLSs that are located in northern hemisphere, with apparent magnitudes brighter than 11, from the 98 candidates in the solar neighborhood provided in a catalogue by \cite{Holberg2013}. The sample are covered from spectroscopic binaries to wide binaries with various masses of WD companions. Table \ref{tab:BasInf} provides the basic information of 21 sample comes from \cite{Holberg2013}.
The successive columns present the primary star name, spectral type of the primary, V magnitude from Tycho satellite, parallax from Hipparcos, Distance in pc (calculated from Hipparcos parallax or from \cite{Holberg2013}), parallax from Gaia, Distance in pc (calculated from Gaia parallax), the mass of the WD in solar mass, binary separation in arcseconds, the estimated semimajor axis ($a$, in au), measured or estimated orbital periods ($p$, in yr), orbital eccentricity, the number of known components in each system, and the name of spectrograph used to take spectra.
%The table with log of observations would give a clear picture rather putting spectrograph column in Table 1
\begin{table*}

\centering
\caption{ Basic information of 21 program stars. Top: sample stars which have abundance analysis in previous literatures; middle: three Ba dwarfs analysed by Paper I; bottom: sample stars analysed for the first time.    }
\label{tab:BasInf}
\scalebox{0.7}{
\begin{tabular}{l|lrrrrrlcrcccr}
\hline\hline
Primary&
Primary&
$P(V\mathrm{t})$  &
$ {\varpi}$    &
D    &
$ {\varpi}$    &
D    &

$M_\mathrm{WD}$    &
$P$  &
$a$  &
$P$    &
 e       &
  Comp  &
 Spectrograph \\
  &
 Type &
   &
(Hip) &
(pc)  &
(Gaia) &
(pc)   &

$({M_\odot})$ &

  (arcsec)  &

  (au)&
  (yr) &
   \\
\hline
HD~218356$^a$	&	K1IV	&	4.91 	&	$5.51\pm 0.23$	&	$181.49\pm7.59$ 	&	-	&	-	&		0.75-1.15$^c$	&	unresolved	&	-	&	0.30	&0.00&	2	& CES \\
HD~202109$^a$	&	G8IIIp	&	3.30 	&	$22.79\pm 0.35$ 	&	$43.88\pm0.67$ 	&	-	&	-	&	0.70-1.10$^c$	&	0.04	&	1.8 	&	17.80	& $0.22\pm 0.03$&	2	& CES \\
HR~5692$^a$	&	G8II-III	&	5.79 	&	$9.15\pm 0.65$ 	&	$109.29\pm7.80$ 	&	-	&	-	&	0.80 	&	orbit	&	2.0 	&	1.71	& $0.34 \pm 0.01$&	2	& CES \\
HD~13611$^a$	&	G6II/III	&	4.45 	&	$8.51\pm 0.51$	&	$117.51\pm7.07$ 	&	-	&	-	&		0.80 	&	<0.16	&	3.8 	&	4.50	&-	&2	& CES \\
HD~26965	&	K0.5V	&	4.51 	&	$200.62\pm 0.23$ 	&	$4.98\pm0.01$ 	&	-	&	-	&	0.51 	&	83.42	&	461.1 	&	8000.00	&-&	3	& CES \\
HR~1608	&	K0IV	&	5.48 	&	$18.53\pm0.84$ 	&	$53.97\pm2.45$ 	&	-	&	-	&	0.35 	&	<0.08	&	<4.8	&	2.47	& $0.30\pm0.06$ &	2	& ARCES \\
\hline
BD+68$^\circ$1027$^b$	&	G5	&	9.78 	&	$12.68\pm0.76$ 	&	$78.86\pm4.74$ 	&	$11.09 \pm 0.22$ 	&	$90.17\pm1.79$ 	&-	&	34.59	&	3027.8 	&	-	& -&	2	& ARCES\\
RE~J0702+129$^b$	&	K0IV/V	&	10.66 	&	-	&	115.00 	&	-	&	-	&	0.57 	&	unresolved	&	NULL	&	-	&-&	3	& ARCES\\
BD+80$^\circ$670$^b$	&	G5V	&	9.16 	&	-	&	40.00 	&	$11.67 \pm 0.72$ 	&	$85.69\pm5.31$ 	&	0.81 	&	18.78	&	833.8 	&	1.87E+04	&-&	2	& ARCES \\
\hline
BD$-$01$^\circ$469	&	K1IV	&	5.62 	&	$14.89\pm0.84$ 	&	$67.16\pm3.80$ 	&	-	&	-	&	0.60 	&	48.43	&	3610.3 	&	-	&-&	2	& CES/ARCES \\
BD$-$00$^\circ$4234	&	K2V	&	9.85 &	$22.13\pm2.01$ 	&	$45.19\pm4.14$ 	&	$23.16\pm0.22$ 	&	$43.18\pm0.41$ 	&	0.49 	&	133.10	&	6676.4 	&	4.86E+05	&-&	2	& ARCES \\
BD$-$01$^\circ$343	&	K0	&	10.12 	&	-	&	65.00 	&	-	&	-	&	-	&	10.75	&	775.6 	&	-	&-&	2	& ARCES \\
BD$-$01$^\circ$407	&	G8/K0V	&	9.23 	&	-	&	37.00 	&	$21.00\pm 0.56$ 	&	$47.63\pm1.27$ 	&	0.87 	&	27.41	&	1125.7	&	2.91E+04	&-&	2	& ARCES \\
BD+39$^\circ$539	&	K3V	&	9.96 	&	-	&	38.00 	&	$24.30\pm 0.25$ 	&	$41.16\pm0.42$ 	&0.79 	&	40.74	&	1718.4 	&	5.60E+04	&-&	2	& ARCES \\
BD$-$07$^\circ$5906	&	G5V	&	9.79 	&	-	&	143.00 	&	$7.50\pm 0.36$ 	&	$133.40\pm6.42$ 	&	0.58 	&	1.01	&	58.9 	&	376.29	&-&	3	& ARCES \\
BD+33$^\circ$2834	&	F8V	&	8.66 	&	$14.35\pm 0.87$  & $69.694.24$ & $13.98\pm 0.23$ 	&	$71.55\pm1.18$ 	&0.54 	&	35.18	&	2721.4 	&	1.11E+05	&-&	2	& ARCES \\
BD+13$^\circ$99	&	G8V	&	9.90 	&	$14.38\pm 1.44$ 	&	$69.54\pm7.03$ 	&	$17.41\pm 0.27$ 	&	$57.43\pm0.89$ 	&0.60 	&	71.77	&	5539.9 	&	-	&-&	2& ARCES	\\
BD+71$^\circ$380	&	G2V	&	9.41 	&	$12.27\pm 1.37$ 	&	$81.50\pm9.21$ 	&	$13.04\pm 0.38$ 	&	$76.66\pm2.23$ 	&-	&	30.00	&	2714.0 	&	-	&-&	2& ARCES	\\
BD+20$^\circ$5125	&	K5V	&	10.18 	&	$20.30\pm 1.40$ 	&	$49.26\pm3.41$ 	&	$18.76\pm 0.30$ 	&	$53.30\pm0.85$ 	&-	&	83.04	&	4540.5 	&	-	&-&	2	& ARCES \\
BD+30$^\circ$2592	&	K0IV	&	9.83 	&	$16.51\pm 1.66$ 	&	$60.57\pm6.15$ 	&	$15.48\pm 0.25$ 	&	$64.61\pm1.04$ 	&0.58 	&	22.95	&	1542.7 	&	-	&-&	2	& ARCES \\
HD~39570	&	F8V	&	7.82 	&	$18.68\pm 0.81$ 	&	$53.53\pm2.33$ 	&	-	&	-	&0.91 	&	90.12	&	5354.8 	&	2.73E+05	&-&	2	& ARCES \\

\hline
\multicolumn{12}{l}{$a$ Ba giants; $b$ Ba dwarfs; $c$ \cite{Merle2016}; }
\end{tabular}
}
\end{table*}

\section{Observations and Data reduction}
\begin{figure}

	\includegraphics[width=\columnwidth]{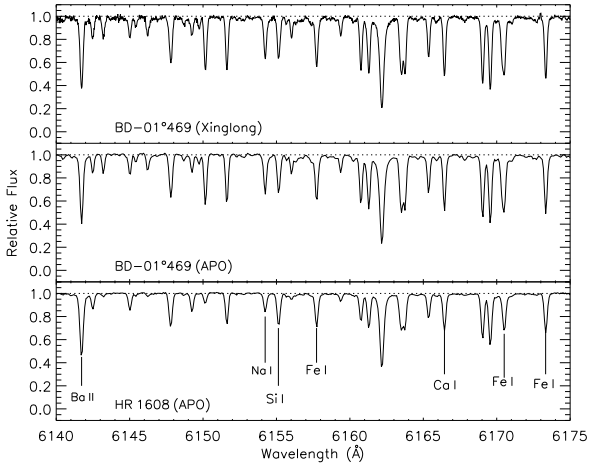}
    \caption{Sample spectra of the normalization of the continuum in the wavelength region from 6140 to 6175\,{\AA}.}
    \label{fig:spec}

\end{figure}

The observations of entire sample were carried out using two different telescopes: i) spectra of 16 stars were obtained with the ARC Echelle Spectrograph (ARCES) attached to the 3.5\,m telescope at the Apache Point Observatory (APO), during three runs of October, November 2014 and February 2015. The ARCES spectral resolving power is $R\sim31\,500\,$ with wavelength coverage from 4\,400\,{\AA} to 10\,000\,{\AA};
ii) spectra of six stars were taken using the Coud\'{e} Echelle Spectrograph (CES) attached to the 2.16\,m telescope operated by National Astronomical Observatories (Xinglong, China), during March and September  2015. The spectra have resolution $R\sim 50\,000\,$ and cover the wavelength range 4\,000\,{\AA} to 9\,000\,{\AA}. For all samples, the exposure time was chosen in order to obtain a S/N of at least 100 over the entire spectral range. In addition, the solar spectra obtained from the two telescopes were used to perform differential abundance analysis. The raw spectra were processed in a standard procedure following Paper I.
%The spectra were reduced using IDL programs consisting of the following standard steps: background subtraction, flat-fielding, order identification and extraction, wavelength calibration, correction of radial velocity and spectrum rectification.
Figure \ref{fig:spec} present the spectra of a dwarf star and a giant star showing main features of absorption in the wavelength region from 6140\,{\AA} to 6175\,{\AA}.

\begin{figure}

	\includegraphics[width=\columnwidth]{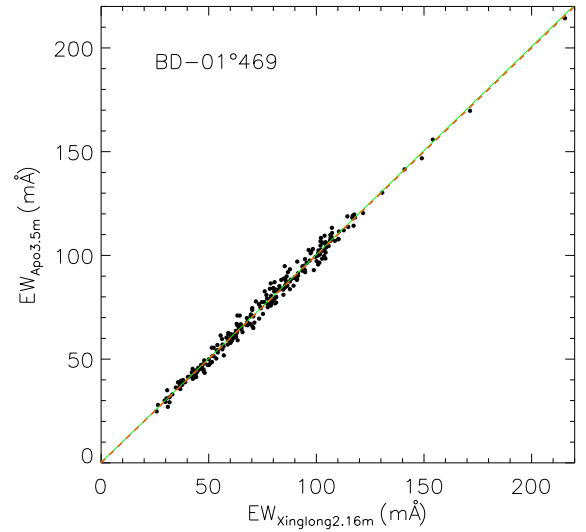}
    \caption{A comparison of equivalent widths (m{\AA}) obtained by Xinglong (2.16 m) with APO (3.5 m) for star BD$-$01$^\circ$469 in common.
    The solid line is a linear fit to the points, whereas the dotted line is the one-to-one relation.}
    \label{fig:ew}

\end{figure}

To check the consistency, the spectra of BD$-$01\,469 taken from two observational set-ups are used to measure EWs of common lines, and are compared (See Figure \ref{fig:ew}). Note, the spectra observed in two set-ups showing good consistency. The systematic difference between the data of two sets is small and a linear regression was obtained, $EW_\mathrm{3.5m} = 1.002 (\pm0.006) EW_\mathrm{2.16m} + 0.162 (\pm0.523) ${(m\AA)}. The standard deviation is about 2.8 m{\AA}.

\section{Analysis and Results}

\subsection{Determination of atmospheric parameters}
%Atmospheric parameters, effective temperature ($T_{\rm eff}$), surface gravity ($\log g$), metallity ([Fe/H]) and microturbulence ($\xi_{t}$), are determined in similar way that was described in Paper I.
Atmospheric parameters like effective temperature ($T_{\rm eff}$), surface gravity ($\log g$), metallicity ([Fe/H]) and microturbulence ($\xi_{t}$), are determined in a similar manner that was described in Paper I.
 The $T_{\rm eff}$ was obtained by imposing excitation equilibrium, i.e, all Fe\,{\sc i} lines with different excitation potentials provide the same abundance. The microturbulent velocity ($\xi_{t}$) was determined by forcing that iron abundances from Fe\,{\sc i} lines of different strengths show no dependence on their equivalent widths (EWs). A linear fit was searched where the angular coefficient is close to zero, and the uncertainty in this coefficient indicates the uncertainty of the $T_{\rm eff}$(spec) and $\xi_{t}$. To estimate 1$\sigma$ uncertainty of these parameters we change their values, respectively, in the plots of Fe\,{\sc i} abundance against excitation potential and EWs, until the angular coefficient of the linear fit match their own uncertainty. The errors of $T_{\rm eff}$(spec) and $\xi_{t}$ estimated by this method are listed in Table 2, and the typical errors are about $\pm$50 K and 0.15 Km$s^{-1}$, respectively.

For comparison, photometric temperatures are derived from $V-K$ colour and empirical calibration relations given by \cite{Alonso1996, Alonso1999, Alonso2001}. Two error sources have been taken into account. One is from the errors of Ks, collected from the Two Micron All-Sky Survey \citep{2003Cutri}, and the other is the measured errors estimated by \cite{Alonso1996,Alonso1999}(see in Table 2).

\begin{figure}%Fig 4
\begin{center}

	% To include a figure from a file named example.* Fig 3
	% Allowable file formats are eps or ps if compiling using latex
	% or pdf, png, jpg if compiling using pdflatex
	\includegraphics[width=\columnwidth]{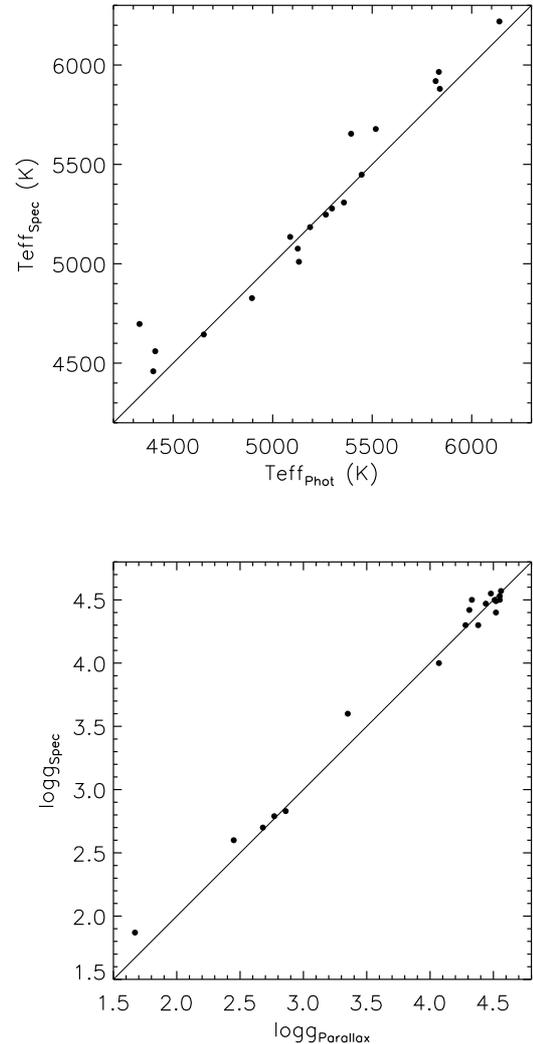}
    \caption{In the top panel, we show the comparison of effective temperatures derived from the photometric colour index $V-K$ and excitation equilibrium methods. In the bottom panel, the comparison of surface gravities obtained by Gaia or Hipparcos parallaxes (X axis)
           and ionization balance of Fe\,{\sc i} and Fe\,{\sc ii} lines (Y axis) are presented.}
\label{tefflog}
\end{center}
\end{figure}

For the stars with available parallaxes, the effective temperatures derived from two methods are compared and shown in Figure \ref{tefflog}. For BD$-$01$^\circ$469 and BD$-$07$^\circ$5906, the difference in temperatures between two methods are as high as 366 K and 260 K. For BD$-$01$^\circ$469, such large error in temperature is probably propagated from error in $K_\mathrm{s}$ band magnitude of 0.234 mag, which is estimated to be 170 K. For BD$-$07$^\circ$5906, the cause for the big difference could be its second companion, which may influence its V or K magnitude. The mean difference between the two methods for other sample stars are
$\langle T_{\rm eff}(spec)-T_{\rm eff}(V-K)\rangle$, is $22\pm79\,\mathrm{K}$. Table \ref{para} list the effective temperatures derived from two methods, and we have adopted the $T_{\rm eff}(spec)$ for abundance analysis.

For stars with available parallaxes, we also derived $\log g$ using standard relation involves mass, temperature, and bolometric flux:
\begin{equation}
\label{math:logg}
\log{g} = \log{g_\odot} +
          \log\left(\frac{M}{M_\odot}\right) +
          4\log\left(\frac{T_\mathrm{eff}}{T_{\mathrm{eff}\odot}}\right) +
          0.4(M_{\mathrm{bol}}-M_{\mathrm{bol}\odot})
\end{equation}

\begin{equation}
\label{math:Mbol}
    M_{\mathrm{bol}}=V_\mathrm{mag}+BC+5\log{\varpi}+5-A_{\mathrm{V}}
\end{equation}

 where M is the stellar mass, which is estimated using an interpolator of the evolutionary tracks of \cite{YonseiYale2003}. The bolometric corrections were calculated using the relation given by \cite{Alonso1995,Alonso1999}.
  %$V_\mathrm{mag}$, $\pi$,$BC$, and $M_\mathrm{bol}$ represent apparent magnitude, parallax, bolometric correction, and absolute bolometric magnitude, respectively.( Paper I )
  For stars with both Gaia and Hipparcos parallaxes, we adopted Gaia parallaxes due to their higher accuracy. From Table 1, we can see that the accuracy of parallaxes in our program stars are high with errors less than 7$\%$. According to equations (\ref{math:logg}) and (\ref{math:Mbol}), this corresponds to an error of ~0.06 dex in $\log g$. Further contributions to the error come from the uncertainty in $V_\mathrm{mag}$, effective temperature and parallax, for which we add 0.04 dex. Finally, the total error estimated for surface gravity is about 0.1 dex.

For comparison, we also calculated the $\log g$ using ionization equilibrium method wherein the value of $\log g$ is obtained by forcing the Fe\,{\sc i} and Fe\,{\sc ii} lines to yield the same iron abundance (See Figure \ref{tefflog}). Table \ref{para} list the surface gravities derived from both methods. The difference of surface gravities between the two methods is small with a mean value of $0.03\pm0.10\,\mathrm{dex}$. For star BD$-$01$^\circ$343, which has no parallaxes from Hipparcos or Gaia, we adopted $\log g$ (Spec) for further analysis. For the rest  of our sample, we adopted $\log g$ (Parallax).

%The $\log g$ (Spec) is often affected by several uncertainties, such as non-local thermodynamic equilibrium (NLTE) effects and only a few available Fe\,{\sc ii} lines, etc. \citep{2007Liu}. Therefore, the error is often as high as 0.2-0.3 dex. For star BD$-$01\,343, there are 13 Fe\,{\sc ii} lines and the dispersion is within 0.08. So, we estimate the errors in surface gravity for our sample are no more than 0.2 dex.}

The initial metallicity values for most of our program stars were set to [Fe/H]= 0.0, and this value will not affect the final result. For the stars with available metallicities in literature, we have taken them as initial metallicities. The final results was adopted by iterating the whole process of determining the atmospheric parameters $T_{\rm eff}$, $\log g$, [Fe/H], and $\xi_{t}$ until they were consistent, and are given in Table \ref{para}. The error estimated in [Fe/H] is $\pm$0.1 dex, which was calculated as described in \cite{1996Ryan}.

\begin{table*}
\centering
\caption{The basic stellar parameters for 21 sample stars.    }
\label{para}
\begin{tabular}{l|ccrclcccrc}
\hline\hline
Primary  &
V-K       &
E(V-K)    &
$M_\mathrm{v}$    &
$M_\mathrm{p}/{M_\odot}$    &
$T_\mathrm{eff}$  &
$T_\mathrm{eff}$  &
$\log g$       &
$\log g$        &
$\mathrm{[Fe/H]}$  &
$\xi_\mathrm{t}$
 \\
 &
 &
 &
 &
 &
 (V-K)&
 (Spec)&
 (Parallax)&
 (Spec)  \\

\hline
HD~218356	&	2.99 	&	0.25 	&	-1.53 	&	2.50 	&	$	4400	\pm	76	$	&	$	4459	\pm	74	$	&	1.67	&	1.87 	&	-0.43	&	$	1.9 	\pm	0.09 	$	\\
HD~202109	&	2.03 	&	0.04 	&	-0.02 	&	3.20 	&	$	5132	\pm	117	$	&	$	5010	\pm	50	$	&	2.68	&	2.70 	&	-0.01	&	$	1.7 	\pm	0.05 	$	\\
HR~5692	&	2.07 	&	0.08 	&	0.50 	&	2.80 	&	$	5126	\pm	361	$	&	$	5076	\pm	46	$	&	2.86	&	2.83 	&	0.02	&	$	1.3 	\pm	0.06 	$	\\
HR~1608	&	1.79 	&	0.07 	&	1.73 	&	1.90 	&	$	5447	\pm	336	$	&	$	5448	\pm	45	$	&	3.35	&	3.60 	&	-0.09	&	$	1.3 	\pm	0.05 	$	\\
HD~13611	&	2.04 	&	0.10 	&	-0.99 	&	3.80 	&	$	5188	\pm	119	$	&	$	5184	\pm	50	$	&	2.45	&	2.60 	&	-0.01	&	$	1.5 	\pm	0.05 	$	\\
HD~26965	&	2.00 	&	0.01 	&	5.93 	&	0.85 	&	$	5088	\pm	112	$	&	$	5135	\pm	45	$	&	4.51	&	4.50 	&	-0.38	&	$	0.7 	\pm	0.06 	$	\\

\hline
BD+68$^\circ$1027	&	1.49 	&	0.04 	&	5.01 	&	0.93 	&	$	5819	\pm	32	$	&	$	5919	\pm	55	$	&	4.48	&	4.55 	&	-0.31	&	$	1.0 	\pm	0.10 	$	\\
RE~J0702+129	&	2.06 	&	0.03 	&	5.28 	&	0.93 	&	$	5052	\pm	42	$	&	$	5531	\pm	60	$	&	4.43	&	4.20 	&	-0.06	&	$	1.9 	\pm	0.12 	$	\\
BD+80$^\circ$670	&	1.57 	&	0.13 	&	4.50 	&	1.05 	&	$	5840	\pm	41	$	&	$	5880	\pm	60	$	&	4.33	&	4.50 	&	0.13	&	$	1.6 	\pm	0.10 	$	\\

\hline
BD$-$01$^\circ$469	&	2.93 	&	0.09 	&	1.48 	&	1.40 	&	$	4331	\pm	170	$	&	$	4697	\pm	55	$	&	2.77	&	2.79 	&	-0.09	&	$	1.2 	\pm	0.06 	$	\\
BD$-$00$^\circ$4234	&	2.66 	&	0.02 	&	6.57 	&	0.56 	&	$	4410	\pm	44	$	&	$	4560	\pm	56	$	&	4.38	&	4.30 	&	-0.96	&	$	0.3 	\pm	0.40 	$	\\
BD$-$01$^\circ$343	&	1.86 	&	0.02 	&	-	&	-	&	$	5292	\pm	46	$	 &	$	5260	\pm	70	$	&	-	&	4.30 	&	0.36	&	$	0.9 	\pm	0.15 	$	\\
BD$-$1$^\circ$407	&	1.88 	&	0.06 	&	5.54 	&	0.83 	&	$	5298	\pm	47	$	&	$	5278	\pm	55	$	&	4.52	&	4.40 	&	-0.12	&	$	0.8 	\pm	0.12 	$	\\
BD+39$^\circ$0539	&	2.48 	&	0.01 	&	6.77 	&	0.78 	&	$	4654	\pm	40	$	&	$	4644	\pm	54	$	&	4.56	&	4.57 	&	0.01	&	$	0.4 	\pm	0.25 	$	\\
BD$-$7$^\circ$5906	&	1.80 	&	0.07 	&	4.09 	&	0.92 	&	$	5394	\pm	48	$	&	$	5654	\pm	44	$	&	4.07	&	4.00 	&	-0.13	&	$	0.9 	\pm	0.10 	$	\\
BD+33$^\circ$2834	&	1.28 	&	0.02 	&	4.37 	&	1.17 	&	$	6139	\pm	40	$	&	$	6219	\pm	65	$	&	4.44	&	4.47 	&	-0.06	&	$	1.2 	\pm	0.08 	$	\\
BD+13$^\circ$99	&	1.92 	&	0.08 	&	6.00 	&	0.73 	&	$	5267	\pm	51	$	&	$	5247	\pm	50	$	&	4.55	&	4.53 	&	-0.38	&	$	0.6 	\pm	0.15 	$	\\
BD+71$^\circ$380	&	1.68 	&	0.05 	&	4.92 	&	0.73 	&	$	5518	\pm	45	$	&	$	5678	\pm	45	$	&	4.31	&	4.42 	&	-0.41	&	$	0.7 	\pm	0.13 	$	\\
BD+20$^\circ$5125	&	2.21 	&	0.02 	&	6.44 	&	0.79 	&	$	4896	\pm	41	$	&	$	4827	\pm	60	$	&	4.55	&	4.50 	&	-0.14	&	$	0.3 	\pm	0.30 	$	\\
BD+30$^\circ$2592	&	1.79 	&	0.02 	&	5.67 	&	0.89 	&	$	5358	\pm	43	$	&	$	5308	\pm	50	$	&	4.52	&	4.49 	&	-0.07	&	$	0.9 	\pm	0.13 	$	\\
HD~39570	&	1.48 	&	0.03 	&	4.11 	&	1.25 	&	$	5835	\pm	39	$	&	$	5965	\pm	35	$	&	4.28	&	4.30 	&	0.03	&	$	1.3 	\pm	0.05 	$	\\

\hline

\end{tabular}
\end{table*}

\subsection{Atomic data, Abundances, and  Error analysis}

 Atomic data for various elements used in this study are the same as those given in paper I, and we added Sulphur (S\,{\sc i}) lines from \cite{Melendez2014}. We have performed the abundance analysis using EWs, and a grid of plane-parallel and local thermodynamic equilibrium models provided by \cite{1993Kur}. The ABONTEST8 program supplied by Dr. P. Magain was used to calculate the theoretical line EWs, and elemental abundances were calculated by requiring that the calculated equivalent width from the model should match the observed value (see paper I for details). Differential abundances ([X/Fe]) are obtained relative to solar values, wherein we adopted $T_{\rm eff}$ = 5780 K, $\log g$ = 4.44, and $\xi_{t}$ = 0.9 km\,s$^{-1}$ to derive solar abundances. The atomic line data and corresponding EWs, and abundances for our program stars are listed in Table \ref{linelist}. The complete table is available in electronic form.

\begin{table} \small
\caption{Atomic line data and measured EWs and absolute
abundances for our program stars.
    }
\label{linelist}
\centering
\scalebox{0.8}{
\begin{tabular}{c|llcccrr}
\hline\hline
Star   &
Ele  &
Ion &
Wavelength  &
$\log gf$           &
$E_\mathrm{low}$   &
EW  &
Abun \\
    &
    &
    &
[{\AA}] &    & [eV] & (m{\AA}) & (dex)  \\
 \hline
BD$-$01$^\circ$469    & C  & 1   & 5380.337& 7.680  & $-$1.570      & 13.4  & 8.398   \\
           & Na & 1  & 6160.750 &  2.100 &  $-$1.240    & 101.0   & 6.326  \\
           & Na & 1  & 6154.230 &  2.100  & $-$1.510      & 86.5 &  6.375  \\
           &Mg  & 1   & 7691.550 & 5.750  & $-$1.000     & 112.1 &  7.699 \\
           &Mg  & 1   & 7387.690 & 5.750  & $-$1.250     &  97.0 &   7.773 \\
           &Mg  & 1   & 8717.830 & 5.930  & $-$1.050     &   98.2 &   7.735 \\
           &Mg  & 1   & 8712.690 & 5.930  & $-$1.310     &  77.5 &   7.769 \\

\hline
\end{tabular}
}
\end{table}

The uncertainties in abundance measurements are estimated by considering the uncertainties in stellar parameters and the measurement errors of EWs.
Table \ref{tbl:error1} and Table \ref{tbl:error2} list the abundance differences when changing the effective temperature by 50 K, the surface gravity by 0.1 dex, the iron abundance by 0.1 dex, and the microturbulent velocity by 0.15 $ \,\mathrm{km\,s^{-1}}$. $\sigma_{stat}$ represents the statistical uncertainty which is line-to-line dispersion divided by $\sqrt{N}$, where N is the number of lines used for a given element. The total error was calculated by taking square root of quadratic sum of the errors associated to all factors, and are given in Col. 8 of Table \ref{tbl:error1} \& \ref{tbl:error2}. These computations for two stars with different characteristics, giant (close binaries) and dwarf (wide binaries), are given in Table \ref{tbl:error1} and Table \ref{tbl:error2}, respectively.

\begin{table} \small
\caption{Abundance errors of giant HR~1608.}
\label{tbl:error1}
\centering
\scalebox{0.7}{
\begin{tabular}{l|rrrrrrr}
\hline\hline
$\Delta [\mathrm{X/H}]$  &
%$\sigma_\mathrm{EW}/\sqrt{N}$   &
N    &
$\sigma_{stat}$ &

$\Delta T_\mathrm{eff}$  &
$\Delta\log g$           &
$\Delta\mathrm{[Fe/H]}$  &
$\Delta\xi_\mathrm{t}$   &
$\sigma_\mathrm{Total}$ \\
    &
    &
    &
\small ($+50\mathrm{K}$)    &
 ($+0.1$) &
 ($+0.1$) &
($+0.15\,\mathrm{km\,s^{-1}}$)    &
 \\

\hline
$\Delta\mathrm{[	C/H 	]}$ &	1	&	-	& $	-0.04 	$ & $	0.03 	$ & $	-0.01 	$ & $	0.00 	$ &	0.05 	\\
$\Delta\mathrm{[	Na/H	]}$ &	3	&	0.06	& $	0.03 	$ & $	-0.01 	$ & $	0.00 	$ & $	-0.02 	$ &	0.07 	\\
$\Delta\mathrm{[	Mg/H	]}$ &	5	&	0.02	& $	0.02 	$ & $	-0.02 	$ & $	0.01 	$ & $	-0.02 	$ &	0.04 	\\
$\Delta\mathrm{[	Al/H	]}$ &	6	&	0.03	& $	0.02 	$ & $	-0.02 	$ & $	0.00 	$ & $	-0.01 	$ &	0.04 	\\
$\Delta\mathrm{[	Si/H	]}$ &	23	&	0.02	& $	0.01 	$ & $	0.00 	$ & $	0.01 	$ & $	-0.02 	$ &	0.03 	\\
$\Delta\mathrm{[	S/H  	]}$ &	3	&	0.03	& $	-0.02 	$ & $	0.03 	$ & $	0.00 	$ & $	-0.01 	$ &	0.05 	\\
$\Delta\mathrm{[	K/H  	]}$ &	1	&	-	& $	0.05 	$ & $	-0.04 	$ & $	0.03 	$ & $	-0.05 	$ &	0.09 	\\
$\Delta\mathrm{[	Ca/H	]}$ &	16	&	0.02	& $	0.04 	$ & $	-0.02 	$ & $	0.00 	$ & $	-0.05 	$ &	0.07 	\\
$\Delta\mathrm{[	ScI/H	]}$ &	3	&	0.01	& $	0.06 	$ & $	0.00 	$ & $	0.01 	$ & $	-0.01 	$ &	0.06 	\\
$\Delta\mathrm{[	ScII/H	]}$ &	5	&	0.01	& $	0.00 	$ & $	0.04 	$ & $	0.03 	$ & $	-0.04 	$ &	0.06 	\\
$\Delta\mathrm{[	TiI/H	]}$ &	22	&	0.02	& $	0.06 	$ & $	-0.01 	$ & $	0.01 	$ & $	-0.05 	$ &	0.08 	\\
$\Delta\mathrm{[	TiII/H	]}$ &	7	&	0.03	& $	0.00 	$ & $	0.04 	$ & $	0.03 	$ & $	-0.05 	$ &	0.07 	\\
$\Delta\mathrm{[	V/H	]}$ &	7	&	0.03	& $	0.06 	$ & $	0.00 	$ & $	0.01 	$ & $	-0.02 	$ &	0.07 	\\
$\Delta\mathrm{[	CrI/H	]}$ &	9	&	0.01	& $	0.05 	$ & $	-0.01 	$ & $	0.01 	$ & $	-0.03 	$ &	0.06 	\\
$\Delta\mathrm{[	CrII/H	]}$ &	5	&	0.03	& $	-0.02 	$ & $	0.04 	$ & $	0.02 	$ & $	-0.04 	$ &	0.07 	\\
$\Delta\mathrm{[	Mn/H	]}$ &	3	&	0.03	& $	0.05 	$ & $	-0.02 	$ & $	0.01 	$ & $	-0.07 	$ &	0.09 	\\
$\Delta\mathrm{[	FeI/H	]}$ &	172	&	0.01	& $	0.04 	$ & $	-0.01 	$ & $	0.00 	$ & $	-0.05 	$ &	0.06 	\\
$\Delta\mathrm{[	FeII/H	]}$ &	19	&	0.02	& $	-0.02 	$ & $	0.04 	$ & $	0.03 	$ & $	-0.05 	$ &	0.07 	\\
$\Delta\mathrm{[	Co/H	]}$ &	8	&	0.04	& $	0.05 	$ & $	0.00 	$ & $	0.01 	$ & $	-0.02 	$ &	0.07 	\\
$\Delta\mathrm{[	Ni/H	]}$ &	42	&	0.01	& $	0.04 	$ & $	0.00 	$ & $	0.01 	$ & $	-0.04 	$ &	0.05 	\\
$\Delta\mathrm{[	Cu/H	]}$ &	4	&	0.05	& $	0.04 	$ & $	-0.01 	$ & $	0.01 	$ & $	-0.07 	$ &	0.10 	\\
$\Delta\mathrm{[	Sr/H	]}$ &	1	&	-	& $	0.06 	$ & $	-0.01 	$ & $	0.01 	$ & $	-0.08 	$ &	0.10 	\\
$\Delta\mathrm{[	Y/H	]}$ &	2	&	0.03	& $	0.00 	$ & $	0.04 	$ & $	0.03 	$ & $	-0.07 	$ &	0.09 	\\
$\Delta\mathrm{[	Zr/H	]}$ &	1	&	-	& $	0.00 	$ & $	0.04 	$ & $	0.03 	$ & $	-0.01 	$ &	0.05 	\\
$\Delta\mathrm{[	Ba/H	]}$ &	3	&	0.06	& $	0.02 	$ & $	0.01 	$ & $	0.05 	$ & $	-0.08 	$ &	0.11 	\\
$\Delta\mathrm{[	La/H	]}$ &	3	&	0.04	& $	0.01 	$ & $	0.04 	$ & $	0.04 	$ & $	-0.01 	$ &	0.07 	\\
$\Delta\mathrm{[	Ce/H	]}$ &	2	&	0	& $	0.01 	$ & $	0.04 	$ & $	0.03 	$ & $	-0.05 	$ &	0.07 	\\
$\Delta\mathrm{[	Nd/H	]}$ &	3	&	0.02	& $	0.01 	$ & $	0.04 	$ & $	0.04 	$ & $	-0.02 	$ &	0.06 	\\

\hline
\end{tabular}
}
\end{table}

\begin{table} \small
\caption{Abundance errors of dwarf BD+71$^\circ$380.    }
\label{tbl:error2}
\centering
\scalebox{0.7}{
\begin{tabular}{l|rrrrrrr}
\hline\hline
$\Delta [\mathrm{X/H}]$  &

N    &
$\sigma_{stat}$ &

$\Delta T_\mathrm{eff}$  &
$\Delta\log g$           &
$\Delta\mathrm{[Fe/H]}$  &
$\Delta\xi_\mathrm{t}$   &
$\sigma_\mathrm{Total}$ \\
    &
    &
  &   \small ($+50\mathrm{K}$)    &  ($+0.1$) & ($+0.1$) & ($+0.15\,\mathrm{km\,s^{-1}}$) &   \\
 \hline
$\Delta\mathrm{[	C/H	]}$ &	1	&	-	& $	0.03 	$ & $	0.03 	$ & $	-0.01 	$ & $	0.00 	$ &	0.04 	\\
$\Delta\mathrm{[	Na/H	]}$ &	3	&	0.02 	& $	-0.03 	$ & $	-0.02 	$ & $	0.00 	$ & $	-0.01 	$ &	0.04 	\\
$\Delta\mathrm{[	Mg/H	]}$ &	6	&	0.01 	& $	-0.03 	$ & $	-0.02 	$ & $	0.01 	$ & $	-0.01 	$ &	0.04 	\\
$\Delta\mathrm{[	Al/H	]}$ &	4	&	0.03 	& $	-0.02 	$ & $	-0.01 	$ & $	0.00 	$ & $	0.00 	$ &	0.04 	\\
$\Delta\mathrm{[	Si/H	]}$ &	21	&	0.01 	& $	-0.01 	$ & $	-0.01 	$ & $	0.01 	$ & $	-0.01 	$ &	0.02 	\\
$\Delta\mathrm{[	K/H	]}$ &	1	&	-	& $	-0.05 	$ & $	-0.05 	$ & $	0.02 	$ & $	-0.02 	$ &	0.07 	\\
$\Delta\mathrm{[	Ca/H	]}$ &	11	&	0.02 	& $	-0.04 	$ & $	-0.03 	$ & $	0.01 	$ & $	-0.02 	$ &	0.06 	\\
$\Delta\mathrm{[	Sc/H	]}$ &	4	&	0.03 	& $	-0.01 	$ & $	0.03 	$ & $	0.03 	$ & $	-0.02 	$ &	0.06 	\\
$\Delta\mathrm{[	TiI/H	]}$ &	22	&	0.02 	& $	-0.06 	$ & $	-0.01 	$ & $	0.00 	$ & $	-0.03 	$ &	0.07 	\\
$\Delta\mathrm{[	TiII/H	]}$ &	5	&	0.03 	& $	-0.01 	$ & $	0.03 	$ & $	0.03 	$ & $	-0.02 	$ &	0.06 	\\
$\Delta\mathrm{[	V/H	]}$ &	5	&	0.04 	& $	-0.06 	$ & $	0.00 	$ & $	0.00 	$ & $	-0.01 	$ &	0.07 	\\
$\Delta\mathrm{[	CrI/H	]}$ &	8	&	0.02 	& $	-0.05 	$ & $	-0.02 	$ & $	0.01 	$ & $	-0.02 	$ &	0.05 	\\
$\Delta\mathrm{[	CrII/H	]}$ &	6	&	0.03 	& $	0.01 	$ & $	0.03 	$ & $	0.02 	$ & $	-0.02 	$ &	0.05 	\\
$\Delta\mathrm{[	Mn/H	]}$ &	4	&	0.03 	& $	-0.04 	$ & $	-0.02 	$ & $	0.00 	$ & $	-0.02 	$ &	0.06 	\\
$\Delta\mathrm{[	FeI/H	]}$ &	124	&	0.01 	& $	0.04 	$ & $	-0.01 	$ & $	0.00 	$ & $	-0.03 	$ &	0.05 	\\
$\Delta\mathrm{[	FeII/H	]}$ &	16	&	0.02 	& $	-0.01 	$ & $	0.03 	$ & $	0.03 	$ & $	-0.03 	$ &	0.05 	\\
$\Delta\mathrm{[	Co/H	]}$ &	5	&	0.03 	& $	-0.04 	$ & $	0.00 	$ & $	0.00 	$ & $	-0.01 	$ &	0.05 	\\
$\Delta\mathrm{[	Ni/H	]}$ &	34	&	0.01 	& $	-0.04 	$ & $	-0.01 	$ & $	0.01 	$ & $	-0.02 	$ &	0.04 	\\
$\Delta\mathrm{[	Cu/H	]}$ &	2	&	0.02 	& $	-0.04 	$ & $	-0.02 	$ & $	0.01 	$ & $	-0.03 	$ &	0.06 	\\
$\Delta\mathrm{[	Sr/H	]}$ &	1	&	-	& $	-0.05 	$ & $	-0.01 	$ & $	0.00 	$ & $	-0.03 	$ &	0.06 	\\
$\Delta\mathrm{[	Y/H	]}$ &	3	&	0.03 	& $	-0.01 	$ & $	0.04 	$ & $	0.03 	$ & $	-0.03 	$ &	0.06 	\\
$\Delta\mathrm{[	Ba/H	]}$ &	3	&	0.04 	& $	-0.02 	$ & $	0.01 	$ & $	0.04 	$ & $	-0.05 	$ &	0.08 	\\
$\Delta\mathrm{[	La/H	]}$ &	1	&	-	& $	-0.02 	$ & $	0.04 	$ & $	0.03 	$ & $	-0.01 	$ &	0.05 	\\
$\Delta\mathrm{[	Ce/H	]}$ &	1	&	-	& $	-0.02 	$ & $	0.04 	$ & $	0.04 	$ & $	-0.02 	$ &	0.06 	\\

\hline
\end{tabular}
}
\end{table}

We note that the uncertainties in Table \ref{tbl:error1} \& \ref{tbl:error2} are the uncertainties for absolute abundance ([X/H]). To estimate the uncertainties on the [X/Fe] abundance, the absolute abundances A(Fe) and A(X) were quadratically added and listed in Table 8, and also showed in respective figures.

\subsection{Comparison with previous studies}

Elemental abundances for six stars in our sample have been determined previously in literature, in which four stars (HD~218356, HD~202109, HR~5692 and HR~1608) have been studied by MER16. Atmospheric parameters and abundances of these stars are listed in Table \ref{ComStarPara} \& \ref{ComStarAbu}. From Table \ref{ComStarPara} it is clear that derived stellar parameters for five sample stars in this study are in good agreement with literature. For HD~218356, there lies obvious difference in $T_{\rm eff}$, $\log g$ and $\xi_{t}$ compare to MER16. However, the methodology employed by MER16 to derive atmosphere parameters are similar to ours; excitation potentials equilibrium method for $T_{\rm eff}$ and ionization equilibrium method for $\log g$. MER16's estimates of temperature, surface gravity and microturbulence are 4244 K, 1.30 and 1.55, respectively, for HD~218356, which are systematically lower than our values, 4459, 1.67 and 1.90.
 We compute the $T_{\rm eff}$ and $\log g$ for HD~218356 using two methods, and from Table \ref{para} we can see that they are consistent.
% which indicate our data is reasonable.
Further in HD~218356, significant differences are found in s-process abundances between ours and MER16 estimates, which could have propagated from difference in atmospheric parameters.

The Ba abundance in four common sample are found to be lower than those of MER16, but are in good agreement with \cite{Silva2015} for HD~202109 and \cite{Merle2014} for HR~1608. We notice that the statistical uncertainties in Ba abundances of MER16 are 0.18, 0.25, 0.07 and 0.22, for HD~218356, HD~202109, HR~5692, HR~1608, respectively, which are larger than errors estimated from this work. For HD 202109, [X/Fe] of most elements are consistent among four data sets within uncertainties. Especially, our values are in good agreement with \cite{Silva2015}, and the mean difference is $+0.03\pm0.05\,\mathrm{dex}$.

\section{Discussion}

\begin{figure*}
\begin{center}

	\includegraphics[width=16cm]{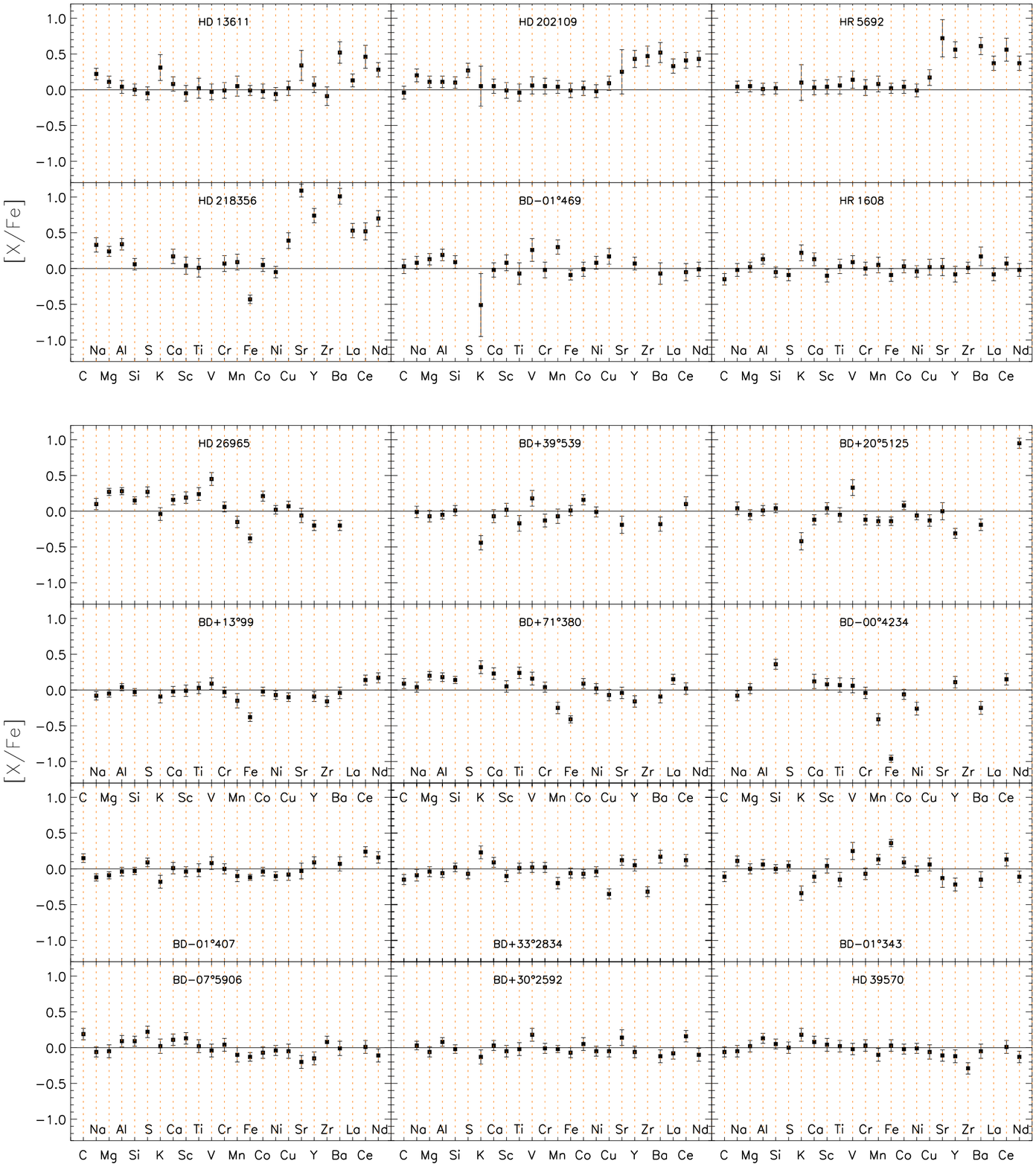}
    \caption{The abundance pattern for 18 sample stars in this work. Each abundance is given with an total error bar which has been listed in Table 8. }
\label{xFe}
\end{center}
\end{figure*}

Among the six giants in our program stars, there are four known mild Ba stars, HD~13611, HD~202109, HR~5692 and HD~218356, which were classified as Ba0.3, Ba1.0, Ba0.3, and Ba2.0, respectively, by \cite{Lu1991}. As shown in figure \ref{xFe}, the s-process elements (Sr, Y, Zr, Ba, La, Ce and Nd) are overabundant in these stars, which are consistent with literature.
Two other giant stars, BD$-$01$^\circ$469 and HR~1608, show no overabundance in s-process elements. Except HD~218356 with [Fe/H] of $-$0.43, the other five giants are distributed in a very narrow range of metallicity: $-0.09 \leq\ [Fe/H] \leq\ 0.02$.

 \subsection{Comparison with field stars}

\begin{figure*}%Fig 6 hd202109 error_compare
\begin{center}

	\includegraphics[width=17cm]{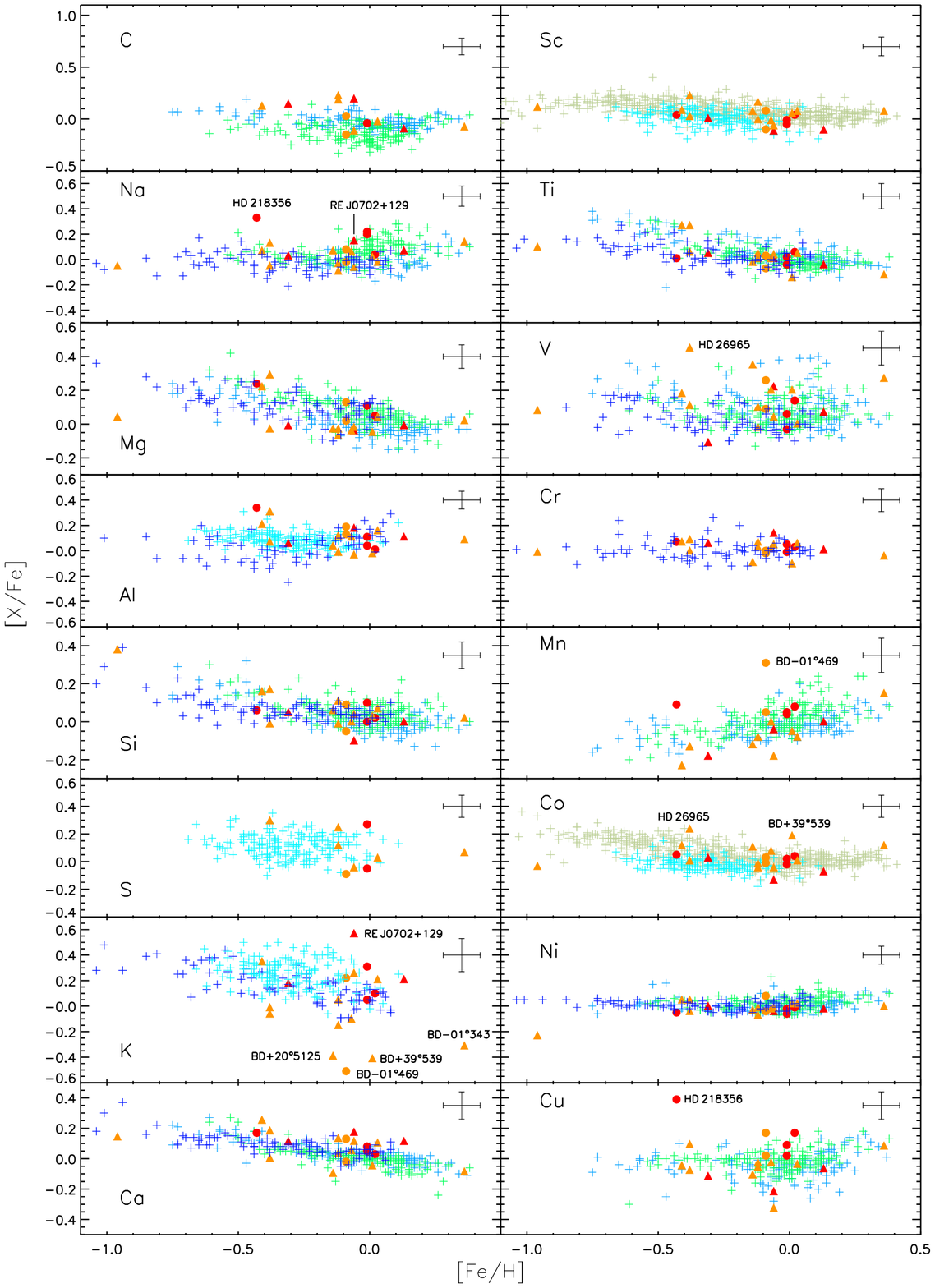}
    \caption[Cap for listoffigures]{Abundance ratio [X/Fe] versus [Fe/H]. Red filled circles: Ba giants in this work; red filled triangles: Ba dwarfs in Paper I; orange filled circles and triangles: normal giant and dwarf stars in this work; blue and green crosses: dwarfs and giants given by \cite{Silva2015}; purple crosses: 90 F and G disk dwarfs given by \cite{chen2000}; fluorescent blue crosses: 181 F and G dwarfs given by \cite{Reddy2003};  Army-green crosses: F and G dwarfs in the solar neighborhood provided by \cite{2015Battistin}. An error bar with the average value of total errors of [X/Fe] is shown in the top right-hand corner of each panel.}
\label{element}
\end{center}
\end{figure*}

\begin{figure*}%Fig 6 hd202109 error_compare
\begin{center}

	\includegraphics[width=17cm]{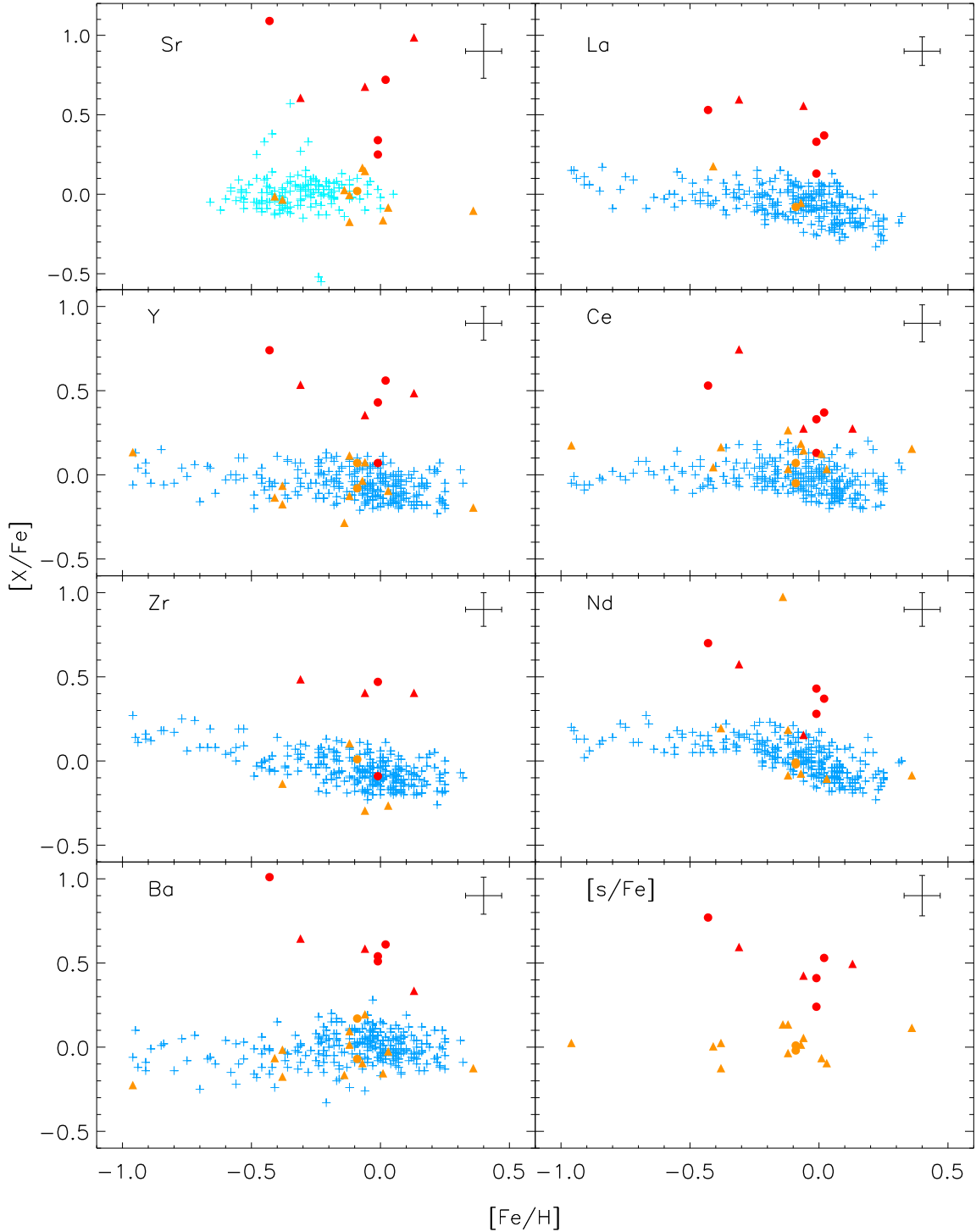}
    \caption[Cap for listoffigures]{Abundance trends of [X/Fe] against [Fe/H] for seven heavy elements (Sr, Y, Zr, Ba, La, Ce and Nd) and the mean values of them. Red filled circles and triangles: Ba giants in this work and Ba dwarfs in Paper I; orange filled circles and triangles: normal giant and dwarf stars in this work; blue crosses: 276 FGK dwarfs from \cite{Mishenina2013}; fluorescent blue crosses: 181 F and G dwarfs from \cite{Reddy2003}.  An error bar with the average value of total errors of [X/Fe] is shown in the top right-hand corner of each panel.}
\label{heavy}
\end{center}
\end{figure*}

The abundance pattern of our sample (normal and Ba stars in binary systems) from this work are compared with the literature values of single FGK field stars from previous studies (see Figure \ref{element} \& \ref{heavy}), to search for differences and similarities that could help to shed some light on the origin of these objects. The literature sample for comparison consists of 90 FG disk dwarfs from \cite{chen2000}, 181 FG dwarfs from \cite{Reddy2003}, 309 FGK dwarfs, subgiants and giants from \cite{Silva2015}, 276 FGK dwarfs from \cite{Mishenina2013}, and a large sample of F and G dwarfs from \cite{2015Battistin} (594 stars for Sc and 567 stars for Co). It is evident from Figure \ref{element} that carbon (C), cobalt (Co), sodium (Na), $\alpha$-elements (Mg, Si, S, Ca, Ti), iron-peak elements (Sc, V, Cr, Mn, Ni, Cu), Odd-Z light elements (Al, K) in most of our sample stars follow the trends of field stars. The s-process elements for seven Ba stars in our sample show different degree of enhancements.

 In figure \ref{element} we can see that normal giant stars are underabundant in [C/Fe] and overabundant in [Na/Fe] in the metallicity range of $-0.3<[Fe/H]<0.2$. A generally used explanation for this differences is that the mixing processes brings up C-poor and Na-rich material to the surface of evolved stars.

C abundance in four Ba stars (HD~202109, RE~J0702+129, BD+80$^\circ$670 and BD+68$^\circ$1027) are just like other normal stars following the abundance distribution of field sample stars. We noticed that two Ba stars (RE~J0702+129 and HD~218356) in our sample show slight overabundance of Na (0.12 and 0.33), whereas other stars have values of [Na/Fe] similar to the field stars of same metallicity. \cite{Silva2015} found that some Ba giants in their sample show excess of Na. It is worth mentioning that these two stars show the signatures of being contaminated by their companion and they have not reached the AGB phase to bring up the Na-rich material to the surface through dredge-up process. \cite{Shi2004} argued that large amount of Na  could be produced by AGB stars. Considering the total uncertainties of [Na/Fe] of the two samples, the Na overabundance is not very obvious. Even so, we still speculate that the two stars might have received excess Na from their companion (during AGB phase) through the mass-loss mechanism.

For Ba and normal field stars (giants and dwarfs) in our sample, the abundance trends of $\alpha$ elements (Mg, Si, Ca and Ti) are similar to the dwarfs and giants studied by \cite{chen2000} and \cite{Silva2015}, showing a slight increase in abundances with decreasing the metallicity. The abundance ratio of Ca in Ba dwarfs, RE~J0702+129 and BD+80$^\circ$670, are marginally higher (0.09 and 0.15) than those seen in normal dwarfs.
We could estimate [S/Fe] for only nine of our sample stars, and the results are consistent with \cite{Reddy2003}.
A large scatter was found in Potassium (K) abundance of our sample. We can see that several sample stars are located far away from the abundance pattern of the disk dwarfs. The EWs of K\,{\sc i} line at 7699\,{\AA} for BD$-$01$^\circ$469, BD$-$01$^\circ$343, BD+20$^\circ$5125, RE~J0702+129 and BD+39$^\circ$539 are from 217.0 to 379.6 m{\AA}, which are stronger and very sensitive to damping constants. It maybe partly responsible for the large scatter.

The abundances of iron peak elements (Sc, V, Cr, Ni and Cu) in our sample follow the trend of field stars, and show no trend in the range of metallicities of our sample. However, the abundance trend shows a slight increase in Mn abundance with increasing the metallicity. \cite{1997Pereira} and \cite{1998Pereira} found two Ba-enriched symbiotic stars present remarkable Cu depletion. \cite{Castro1999} analyzed Cu and Ba abundances for seven Ba dwarfs and found the deficiency of Cu in Ba stars compared to normal disc stars of same metallicity range. It is a possible indication that Cu could be a seed to the production of s-process elements. \cite{Smiljanic2007} and \cite{Allen2011} studied the correlation between the neutron-capture elements and iron peak elements, and verified whether they act as neutron seeds or poisons during the operation of the s-process, but did not find any supportive conclusion. In figure \ref{element}, it is clear that for three Ba dwarfs, the Cu, Ni, and Mn abundances are located at the bottom part of abundance distribution of field dwarfs, but still within the trend, which suggests the relation with heavy elements is unlikely.

We compared our [Co/Fe] ratios with those of \cite{Reddy2003} and \cite{2015Battistin}, and found that the trend of our sample is consistent with those of them.

A small but distinct difference was noticed in trends between \cite{chen2000} and \cite{Reddy2003} for [Al/Fe] against [Fe/H]. \cite{Reddy2003} showed [Al/Fe] ratios have smooth decrease with increasing [Fe/H] in the whole range of metallicities for their sample, but the sample of \cite{chen2000} showed a rather steep upturn of [Al/Fe] beginning at [Fe/H] $\simeq$ $-$ 0.4. Our results are in good agreement with \cite{Reddy2003} and there is no obvious difference between the giants and dwarfs in our sample.

In figure \ref{heavy}, we can see that seven Ba stars in our sample show overabundance of heavy elements when they are compared with the sample of similar metallicity studied by \cite{Mishenina2013}. For the seven Ba stars, the abundances of s-process elements and their [s/Fe] (mean value of Sr, Y, Zr, Ba, La, Ce and Nd) shows anti-correlation with [Fe/H].

\subsection{Ba stars and the masses of WD companions}

The minimum CO core mass at the base of the AGB was predicted to be 0.51 $M_\odot$ (for a star with initial mass 0.9 $M_\odot$) by Eq.(66) of \cite{Hurley2000}. Otherwise, the s-process synthesis would not occur in a star because the progenitor of the WD could not reach to the TP-AGB phase. MER16 analyzed 11 binary systems having WD companions of various masses and showed that the trend of s-process enrichment of their sample supported the prediction except two marginal cases, namely DR\ Dra (0.55 $M_\odot$) and 14\ Aur\ C (0.53 - 0.69 $M_\odot$). For 14\ Aur\ C, MER16 indicated that the absence of s-process enrichments could be this star might have experienced the common-envelope process which does not lead to substantial accretion. For star DR~Dra, the reason is still puzzling.

\begin{figure*}%Fig 6 hd202109 error_compare
\begin{center}

	% To include a figure from a file named example.
	% Allowable file formats are eps or ps if compiling using latex
	% or pdf, png, jpg if compiling using pdflatex
	\includegraphics[width=16cm]{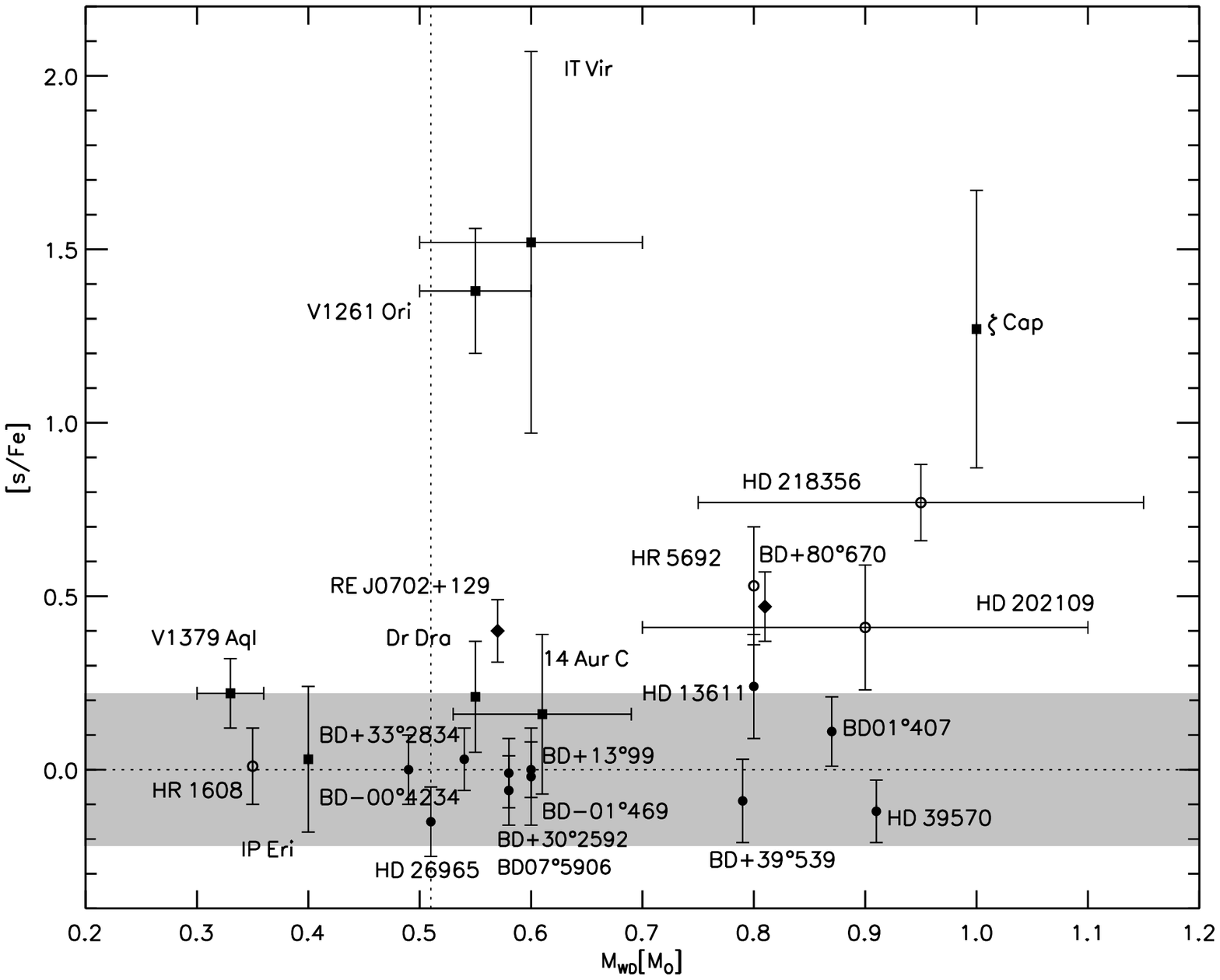}
    \caption{Average s-process abundances relative to Fe as a function of the WD masses. Total error bars are given here. Filled circles: sample stars in this work; filled rhombus: Ba dwarfs from Paper I; filled rectangle: stars from MER16; open circles: common stars in both studies. To guide the eye, the gray zone shows the $\pm0.22$ dex zone without significant enhancement in s-process abundances.}
\label{s_Mwd}
\end{center}
\end{figure*}

There are 17 binary systems whose WD masses are known in our 21 samples (including three Ba dwarfs from Paper I). Except four common stars (HD~218356, HD~202109, HR~5692 and HR~1608) we include other seven sample from MER16 to our sample, and present the average s-process abundances against the masses of WD companions in figure \ref{s_Mwd}. It is clear that four stars whose WD companion masses are less than 0.51 $M_\odot$ do not indeed show presence of s-process enrichment, supporting the prediction of \cite{Hurley2000} and previous observations by \cite{Merle2016}. Interestingly, except star DR\ Dra and 14\ Aur\ C, there are still nine stars whose WD masses qualify for them being polluted by s-process elements, but they are showing absence of s-process enrichment. The largest separation between the seven Ba stars and their WD companions in our sample is about 3028 AU (for BD+68$^\circ$1027), and seven samples whose WD companion masses are larger than 0.51 $M_\odot$ and orbital separations are less than 3028 AU are showing the absence of s-process enrichment.
 Consequently, above analysis indicates a large enough mass of WD with a small enough orbital separation in binaries is not a sufficient condition to form a Ba star. On the positive side, 0.51 $M_\odot$ is probably the threshold to indicate whether the progenitor of a WD has reach the TP-AGB phase.

\begin{figure*}%Fig 6 hd202109 error_compare
\begin{center}

	\includegraphics[width=16cm]{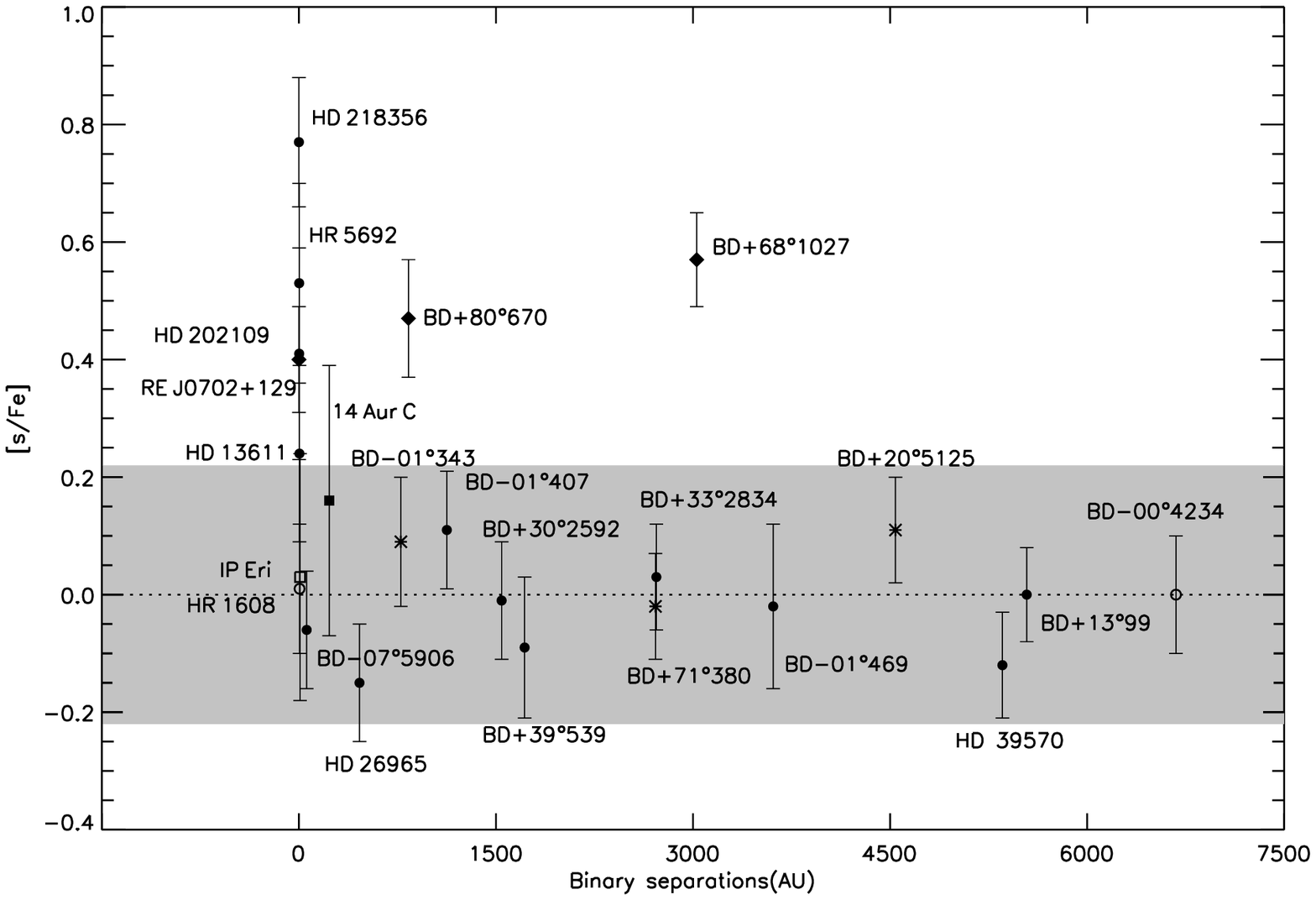}
    \caption{Abundance trends of [s/Fe] against orbital separation (AU). Total error bars are given here.
    Black filled circle: sample stars with WD mass equal or larger than 0.51$M_\odot$ in this work; open circles: sample stars with WD mass less than 0.51$M_\odot$ in this work; stars: sample stars with unknown WD mass in this work; filled rhombus: Ba dwarfs from Paper I; filled rectangles: sample stars with WD mass larger than 0.51$M_\odot$ in MER16's work; open rectangles: sample stars with WD mass less than 0.51$M_\odot$ in MER16's work.
    }
\label{separa}
\end{center}
\end{figure*}

\subsection{Level of s-process enrichment and the orbital separation}

The orbital separation of the binary systems might govern the efficiency of mass transfer from s-process enriched AGB stars, and it has been widely assumed to be one of the important reasons for different levels of s-process enrichment in Ba stars \citep{Han1995, Pols2003}. In this work, we use the estimated semimajor axis (see Table \ref{tab:BasInf}) to stand for the orbital separations. RE~J0702+129 and HD~218356 are unresolved binaries, so we use zero value to represent their orbital separations. We showed [s/Fe] against orbital separations in figure \ref{separa}. Our 21 program stars together with two stars from MER16 (14\ Aur\ C and IP\ Eri), which had been catalogued in \cite{Holberg2013} and provided the semimajor axis, have been presented. We noticed that, except seven Ba stars, there is no apparent overabundances of s-process elements in other 16 stars. For the seven Ba stars, we found no obvious correlation between their [s/Fe] ratios and orbital separations. We can conclude that the binary orbital separation is not the only reason that cause the difference between the mild and the strong Ba stars.

\subsection{Level of s-process enrichment and the metallicity}

Another possible reason that cause the difference between the mild and strong Ba stars is the efficiency of the neutron-capture nucleosynthesis in AGB stars, which is controlled by the metallicity \citep{Jorissen1998, Kappeler2011}. The [hs/ls] ratio has been widely used to measure the s-process efficiency since \cite{Luck1991}. [hs] and [ls] stands for the mean abundance of `heavy' and `light' s-process elements, respectively.

\begin{figure}
\begin{center}

	\includegraphics[width=\columnwidth]{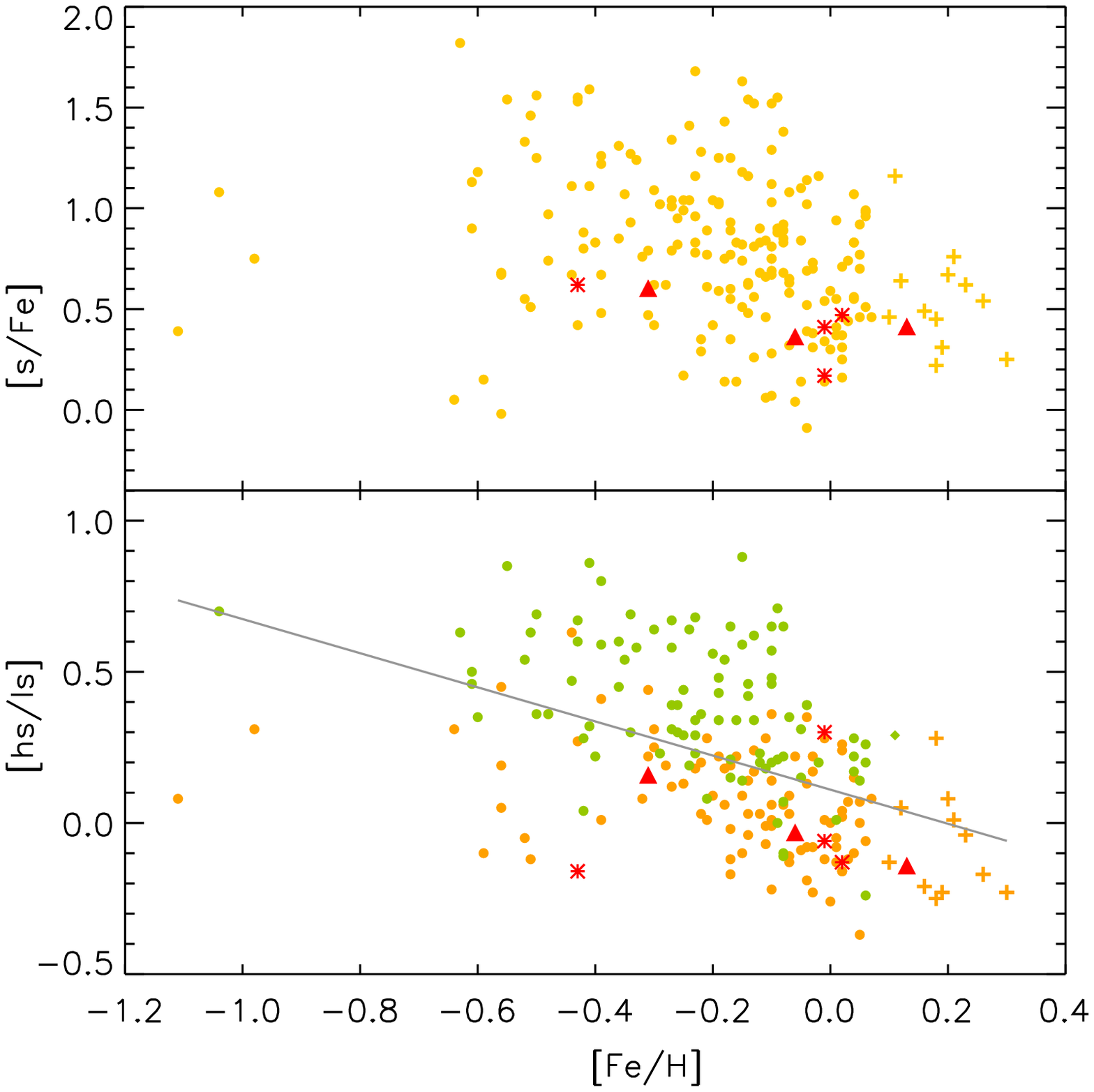}
    \caption[Cap for listoffigures]{The ratio [hs/ls] vs [Fe/H] (red stars: Ba giants in this work; red triangles: Ba dwarfs in Paper I), together with the results from \cite{Castro2016} (filled circles) and \cite{Pereira2011} (crosses). Bottom panel: green color: the sample with [s/Fe] > 0.8; orange color: the sample with [s/Fe] < 0.8. The solid line is a linear fit to all sample stars. For consistency with \cite{Castro2016} and \cite{Pereira2011}, for this Figure, Zr and Y are used for [ls], and La, Ce, and Nd are used for [hs]. s is the mean value of Zr, Y, Ce, Nd and La.}
\label{hs/ls}
\end{center}
\end{figure}

In figure \ref{hs/ls}, we show [hs/ls] ratio and [s/Fe] ratio of our sample along with Ba stars analyzed in \cite{Castro2016} and \cite{Pereira2011}. Different set of elements were used to calculate [hs] and [ls] in literature. Here, to maintain consistency among the three data sets, we adopted Zr and Y to calculate [ls] and La, Ce and Nd to calculate [hs]. 's' is the mean value of Zr, Y, Ce, Nd and La. We found that there is an anti-correlation for [hs/ls] vs. [Fe/H] and [s/Fe] vs. [Fe/H], which is consistent with literature \citep{Smiljanic2007,Pereira2011}.
It has been suggested  in literature that the different level of s-process elemental overabundance is controlled by the metallicity. However, we found  no significant difference in metallicity between strong Ba and mild Ba stars when we put our sample together with the data provided by \cite{Castro2016} and \cite{Pereira2011}. In order to inspect the relation between [hs/ls] ratio and level of n-capture efficiency, in all data sets, the stars with [s/Fe] larger than 0.8 are plotted in green color and represent `strong Ba stars'. The others which are plotted orange color represent `mild Ba stars'. We found no distinct separation between `strong Ba stars' and `mild Ba stars' in the range of $-1.0<[Fe/H]<0.3$. The overlap will be even more obvious if we set the threshold to be 0.6 for 'strong Ba stars'. Even so, we can see that there is a general trend for the `mild Ba stars' to fall below the `strong Ba stars' of the same metallicity.

\subsection{The ratio [hs/ls] and the masses of WD companions}

\begin{figure*}%Fig 6 hd202109 error_compare
\begin{center}

	% To include a figure from a file named example.
	% Allowable file formats are eps or ps if compiling using latex
	% or pdf, png, jpg if compiling using pdflatex
	\includegraphics[width=16cm]{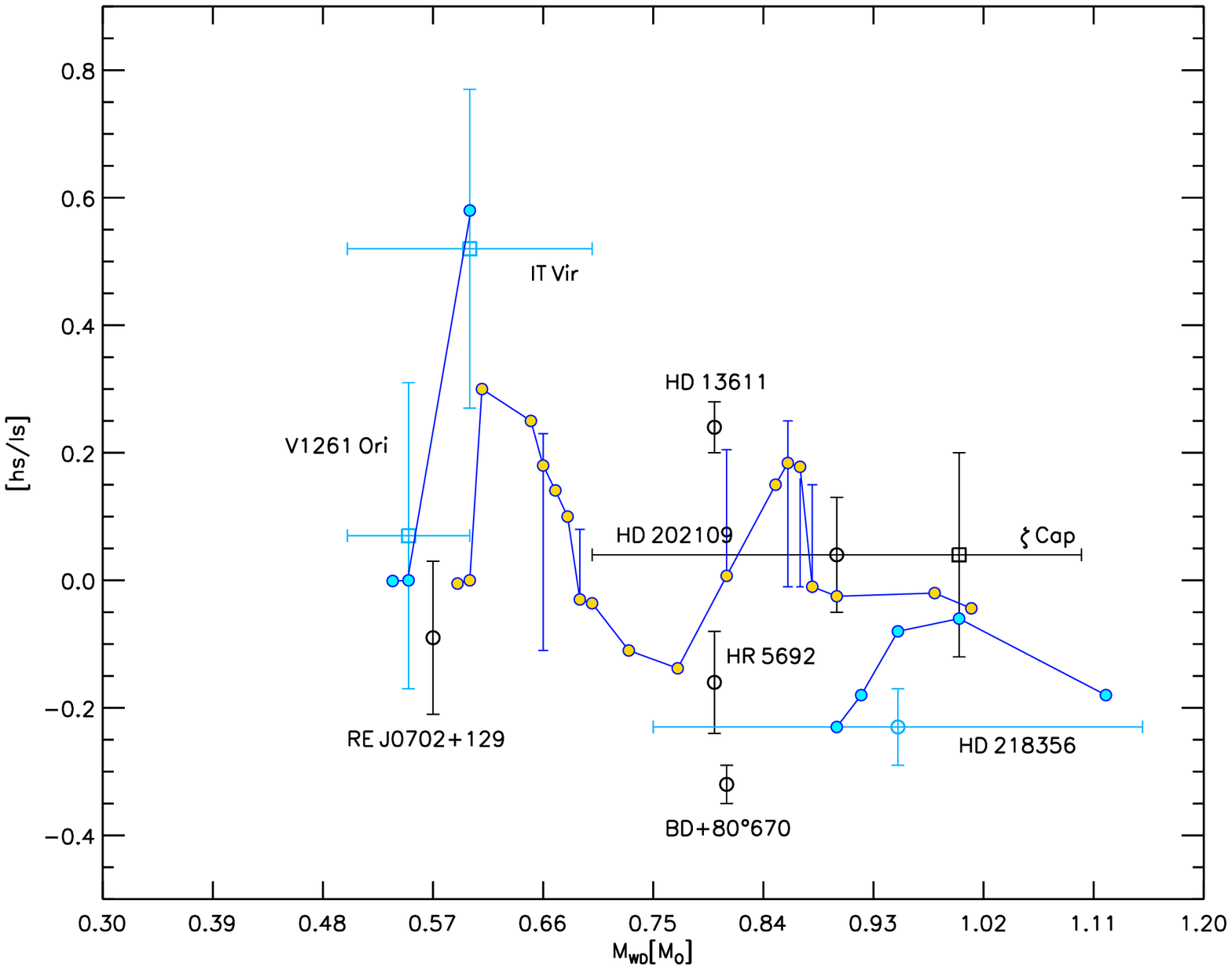}
    \caption[Cap for listoffigures]{ [hs/ls] ratio as a function of WD mass. The yellow filled circles joined by blue lines represent the theoretical models from KL2016 for Z = 0.014. The filled azure circles represent the same models for Z = 0.007. For clarity, the Z=0.014 models are only depicted close to the metal poor stars in our sample which are represented by azure open circles (this work) and rectangles (MER16). Black open rectangles and circles represent stars with solar metallicity from MER16 and this work. Total error bars are given for observed sample. }
\label{hl_Mwd}
\end{center}
\end{figure*}

\cite{2016Karakas} (hereafter KL2016) predicted that low-mass AGB stars produce a higher proportion of [hs] with respect to [ls] elements at solar metallicity. \cite{2017Del} analyzed chemical abundances of 1111 FGK dwarf stars and found that thick disc stars have less Ba with respect to light-s (lower [hs/ls]) elements compare to thin disc stars at a given metallicity. Based on the models of KL2016, they speculated that this phenomenon is probably due to old age of thick disc stars, and at the time of their formation less low-mass AGB stars (which evolve slower than intermediate-mass AGBs) might have contributed to the interstellar medium enrichment, from which they were formed. Ba stars accrete s-process enhancement material from their AGB companion, in due process, the accretion efficiency and pollution factor play a significant role (see paper I for details). After the mass transfer, the abundance pattern of AGB stars will not change but the absolute value of ratio [hs/ls] should be smaller. Even so, in Figure \ref{hl_Mwd}, we still compared [hs/ls] ratio of nine Ba stars for which WD masses are available, with those models from KL2016 (z = 0.014), in order to find whether any relation between them. For three observed stars with low metallicities, V1261\ Ori ([Fe/H] = $-$0.5), IT\ Vir ([Fe/H] = $-$0.5) and HD\ 218356 ([Fe/H] = $-$0.43), we compared them with models (Z = 0.007). The mass of AGB stars (Figure 15 \& 17 from KL2016 ) have been transformed to their core mass based on the models provided by KL2016. From Figure 10, we can see that, overall, the observed values are coincident with the theoretical models. The [hs/ls] of BD+80$^\circ$670 is lower, which might be caused by the contribution of metal rich AGB stars that provides higher abundances of light-s elements (See Figure 10 \& 16 in KL2016). It is worth noting that the s-process elements abundance do not show any enrichment for the models (See Table 4 in KL2016) whose initial masses are 1.25$M_\odot$ (core mass is 0.60$M_\odot$ with Z = 0.014 and 0.62$M_\odot$ with Z = 0.007). However, two observed stars, RE\ J0702+129 and V1261\ Ori, whose WD companion masses are 0.57$M_{\odot}$ and 0.55$M_{\odot}$, respectively, show significant s-process elements enrichment. So, for the minimum CO core mass at the base of the AGB stars, our result are not matching with theoretical models of KL2016, but consistent with the prediction of \cite{Hurley2000}.

\section{Conclusions}
Based on high resolution and high S/N spectra, we carried out a detailed abundance analysis for 18 primary stars of SLSs, in which four known Ba giants, two normal giants, and 12 dwarfs. We determined atmospheric parameters, masses and abundances for 24 elements: C, Na, Mg, Al, Si, S, K, Ca, Sc, Ti, V, Cr, Mn, Fe, Co, Ni, Cu, Sr, Y, Zr, Ba, La, Ce, Nd.
The abundance pattern of Al, $\alpha$ elements, iron-peak elements and odd-Z elements of Ba and normal stars in binaries are similar to those of the field stars with same metallicity. We did not find any difference in the behavior of [X/Fe] between Ba giant and dwarf stars analyzed in this study. Some stars in our sample show enhancement in sodium, compare to the field stars, and the overabundance is probably polluted by WD companions when they are in AGB phase. In previous studies, some iron peak elements have been found to have a relation with the neutron-capture elements and been predicted to act as neutron seeds or poisons during the operation of the s-process. However, we did not find any relation between the neutron-capture elements and iron peak elements.

We found correlation between s-process abundances of primary stars and masses of their WD companions in the binaries, supports the prediction made by \cite{Hurley2000}, wherein s-process synthesis occur during the AGB phase when the WD mass is 0.51 $M_\odot$ or above.
So the primary star may not show s-process elemental enhancement with WD companion of mass less than 0.51 $M_\odot$.
We also found normal s-process abundances in stars with WD mass larger than 0.51 $M_\odot$, which suggests WD mass may not be the only factor to decide the chemical peculiarities in the primary star of SLSs, complicates the present understanding of Ba stars.
We did not find any relation between s-process enrichment and orbital separation of the binaries, suggests the orbital separation may not be the only reason to differentiate mild and strong Ba stars, nor a key factor to form a Ba star or not. Maybe a threshold value need to be found to reject Ba candidates, just like WD mass constraint, and it requires further investigation on theoretical understanding.
[s/Fe] and [hs/ls] ratios of our sample show anti-correlation with metallicity, but we did not find any significant difference in metallicity between strong and mild Ba stars. Strong and mild Ba stars in our sample are small in number, so the threshold 0.8 is used to divide the Ba stars when including the literature sample. The correlation between the [hs/ls] ratio and the level of s-process enhancement is not so clear, suggesting the boundary value might be a key factor. The ratio [hs/ls] as a function of WD mass is consistent with the theoretical models from KL2016.

%\begin{landscape}
\begin{table}
\centering
\caption{Atmosphere parameters for the common sample stars.    }
\label{ComStarPara}
%\scalebox{0.78}{
\begin{tabular}{l|rrrrr}
\hline\hline
Star  &
$T_\mathrm{eff}$       &
$\log g$     &
$\mathrm{[Fe/H]}$   &
$\xi_\mathrm{t}$    &
Ref

 \\

\hline
HD~218356	&	4348	&	1.34 	&	$	-0.58 	$	&	2.19 	&	a	\\
	&	4244	&	1.30 	&	$	-0.45 	$	&	1.55 	&	b	\\
	&	4459	&	1.67 	&	$	-0.43 	$	&	1.90 	&	h	\\
%\hline													
HD~202109	&	5050	&	2.80 	&	$	0.01 	$	&	1.45 	&	c	\\
	&	5129	&	3.02 	&	$	0.08 	$	&	1.74 	&	d	\\
	&	4950	&	2.70 	&	$	-0.03 	$	&	1.43 	&	b	\\
	&	5010	&	2.68 	&	$	-0.01 	$	&	1.70 	&	h	\\
													
HR~5692	&	5032	&	3.00 	&	$	-0.08 	$	&	1.35 	&	b	\\
	&	5076	&	2.86 	&	$	0.02 	$	&	1.30 	&	h	\\
													
HR~1608	&	4960	&	3.30 	&	$	0.09 	$	&	1.50 	&	e	\\
	&	5374	&	3.60 	&	$	-0.18 	$	&	1.03 	&	b	\\
	&	5448	&	3.35 	&	$	-0.09 	$	&	1.30 	&	h	\\

HD~13611	&	5129	&	2.47 	&	$	-0.05 	$	&	1.63 	&	f	\\
	&	5184	&	2.45 	&	$	-0.01 	$	&	1.50 	&	h	\\
													
HD~26965	&	5153	&	4.39 	&	$	-0.31 	$	&	0.36 	&	g	\\
	&	5135	&	4.51 	&	$	-0.38 	$	&	0.70 	&	h	\\

\hline
\multicolumn{6}{l}{$a$ \citealt{2010Bubar}; $b$ \citealt{Merle2016}; }\\
 \multicolumn{6}{l}{$c$ \citealt{Yushchenko2004}; $d$ \citealt{Silva2015}; }\\
 \multicolumn{6}{l}{ $e$ \citealt{Merle2014}; $f$ \citealt{Luck2014};  }\\
 \multicolumn{6}{l}{$g$ \citealt{2012Adibekyan}; $h$ this work}\\
\end{tabular}
%}
\end{table}
%\end{landscape}

\begin{landscape}
\begin{table}
\centering
\caption{Abundance of the common stars. The statistical uncertainties are given here.    }
\label{ComStarAbu}
%\scalebox{0.67}{
\begin{tabular}{l|rrrrrrrrrrrrr}
\hline\hline
Star  &
[C/Fe]  &
[Na/Fe]    &
[Mg/Fe]   &
[Al/Fe]    &
[Si/Fe]  &
[S/Fe] &
[K/Fe]      &
[Ca/Fe]      &
[Sc/Fe]  &
[Ti/Fe] &
[V/Fe]  &
[Cr/Fe]  &
%[Mn/Fe]   &
%[Fe/H]  &
%[Co/Fe] &
%[Ni/Fe] &
%[Cu/Fe] &
%[Sr/Fe] &
%[Y/Fe] &
%[Zr/Fe] &
%[Ba/Fe] &
%[La/Fe] &
%[Ce/Fe] &
%[Nd/Fe] &
Ref
 \\

\hline
HD~218356	&	-	&	$0.53 \pm 0.08$	&	-	&	$0.51 \pm 0.11$	&	-	&	-	&	-	&	-	&	-	&	$0.07 \pm 0.24$	&	-	&	$-0.03 \pm 0.20$  & a	\\
           &	$0.01\pm0.23$	&	-	&	-	&	-	&	-	&	-	&	-	&	-	&	-	&	-	&	-	&	- & b  \\

           &	-	&$	0.33\pm0.10	$&$	0.24\pm0.10	$&$	0.34\pm0.09	$&$	0.06\pm0.10	$&	-	&	-	&$	0.17\pm0.05	$&$	0.04\pm0.09	$&$	0.01\pm0.01	$&	-	&$	0.07\pm0.02$ & h \\

\hline
HD~202109	&	-	&$	0.24\pm0.08	$&$	0.22\pm0.22	$&$	0.12\pm0.12	$&$	-0.05\pm0.13	$&$	0.00\pm0.12	$&	$-$0.17	&$	0.02\pm0.09	$&$	0.07\pm0.15	$&$	0.15\pm0.17	$&$	-0.03\pm0.14	$&$	-0.12\pm0.13	$ & c \\

   &$	-0.09\pm0.06	$&$	0.18\pm0.09	$&$	0.07\pm0.10	$&	-	&$	0.02\pm0.09	$&	-	&	-	&$	-0.01\pm0.12	$&	-	&$	-0.02\pm0.11	$&$	0\pm0.13	$ &	-  & d \\
   &	$-0.22\pm0.13$	&	-	&	-	&	-	&	-	&	-	&	-	&	-	&	-	&	-	&	-	&	- & b \\

  &$	-0.04\pm0.03	$&$	0.20\pm0.06	$&$	0.11\pm0.06	$&$	0.11\pm0.06	$&$	0.10\pm0.09	$&	0.27	&	0.05	&$	0.05\pm0.08	$&$	-0.01\pm0.10	$&$	-0.04\pm0.09	$&$	0.06\pm0.06	$&$	0.05\pm0.07	$ & h\\

																																																\hline
HR~5692	&$	-0.24\pm0.10	$&	-	&	-	&	-	&	-	&	-	&	-	&	-	&	-	&	-	&	-	&	- & b \\
&	-	&$	0.04\pm0.03	$&$	0.05\pm0.05	$&$	0.01\pm0.04	$&$	0.02\pm0.07	$&	-	&	0.10	&$	0.03\pm0.05	$&$	0.04\pm0.06	$&$	0.06\pm0.09	$&$	0.14\pm0.06	$&$	0.03\pm0.08	$& h\\

\hline
HR~1608	&$	0.08\pm0.16	$&$	0.19\pm0.11	$&$	0.18\pm0.10	$&$	0.21\pm0.04	$&$	-0.05\pm0.10	$&	-	&	-	&$	0.19\pm0.04	$&	-	&$	0.11\pm0.15	$&	-	&	- & e \\
       &$	-0.12\pm0.18	$&	-	&	-	&	-	&	-	&	-	&	-	&	-	&	-	&	-	&	-	&	- &b \\
       &	-0.15	&$	-0.02\pm0.06	$&$	0.02\pm0.05	$&$	0.13\pm0.04	$&$	-0.05\pm0.06	$&	-0.09	&	0.22	&$	0.13\pm0.07	$&$	-0.10\pm0.06	$&$	0.03\pm0.11	$&$	0.09\pm0.08	$&$	0.00\pm0.07	$& h\\

\hline

HD~13611	&	$-$0.57	&	0.27	&	$-$0.02	&	0.13	&	0.06	&	0.11	&	-	&	0.02	&	$-$0.06	&	$-$0.05	&	$-$0.08	&	$-$0.01 & f \\

&	-	&$	0.22\pm0.03	$&$	0.11\pm0.06	$&$	0.04\pm0.04	$&$	0.00\pm0.08	$&	-0.05	&	0.31	&$	0.08\pm0.07	$&$	-0.05\pm0.04	$&$	0.02\pm0.08	$&$	-0.03\pm0.01	$&$	-0.01\pm0.09	$& h\\

\hline
HD~26965	&	-	&$	0.16\pm0.01	$&$	0.25\pm0.10	$&$	0.33\pm0.01	$&$	0.14\pm0.03	$&	-	&	-	&$	0.16\pm0.10	$&$	0.17\pm0.13	$&$	0.37\pm0.11	$&$	0.51\pm0.19	$&$	0.07\pm0.03	$ & g \\
&	-	&$	0.10\pm0.10	$&$	0.27\pm0.03	$&$	0.28\pm0.07	$&$	0.15\pm0.06	$&	0.27	&	-0.04	&$	0.16\pm0.06	$&$	0.19\pm0.10	$&$	0.24\pm0.10	$&$	0.43\pm0.07	$&$	0.06\pm0.03	$& h \\

																																																			\hline
%\multicolumn{26}{l}{$a$ \citealt{2010Bubar}; $b$ \citealt{Merle2016}; $c$ \citealt{Yushchenko2004}; $d$ \citealt{Silva2015}; $e$ \citealt{Merle2014}; $f$ \citealt{Luck2014};
% $g$ \citealt{2012Adibekyan}; $h$ this work };\\
\multicolumn{14}{l}{reference: the same as Table \ref{ComStarPara}}\\
\end{tabular}
%}
\end{table}
\end{landscape}

\begin{landscape}
\begin{table}
\centering
\contcaption{Abundance of the common stars. The statistical uncertainties are given here.   }
\label{tab:continued}
%\scalebox{0.9}{
\begin{tabular}{l|rrrrrrrrrrrrrr}
\hline\hline
Star  &
%[C/Fe]  &
%[Na/Fe]    &
%[Mg/Fe]   &
%[Al/Fe]    &
%[Si/Fe]  &
%[S/Fe] &
%[K/Fe]      &
%[Ca/Fe]      &
%[Sc/Fe]  &
%[Ti/Fe] &
%[V/Fe]  &
%[Cr/Fe]
[Mn/Fe]   &
[Fe/H]  &
[Co/Fe] &
[Ni/Fe] &
[Cu/Fe] &
[Sr/Fe] &
[Y/Fe] &
[Zr/Fe] &
[Ba/Fe] &
[La/Fe] &
[Ce/Fe] &
[Nd/Fe] &
Ref
 \\

\hline
HD~218356	&	$-0.04\pm0.09$ &	$-0.58\pm 0.07$	&	-	&	$-0.04\pm0.14$	&	-	&	-	&	-	&	-	&	-	&	-	&	-	&	-	&	a  \\
 &	-&	$-0.45\pm0.12$	&	-	&	-	&	-	&$	0.58\pm0.06	$&$	0.46\pm0.22	$&$	0.31\pm0.07	$&$	1.46\pm0.18	$&$	0.44\pm0.09	$&$	0.16\pm0.09	$&	-	&	b \\
&	0.09	&$	-0.43\pm0.11	$&$	0.05\pm0.09	$&$	-0.05\pm0.09	$&	0.39	&	1.09	&	0.74	&	-	&	1.01	&$	0.53\pm0.01	$&	0.52	&$	0.70\pm0.06	$&	h	\\

																																																\hline
HD~202109 &$	-0.25\pm0.17	$ & $	0.01\pm0.11	$ & $	0.07\pm0.11	$ & $	-0.09\pm0.17	$&	-0.01	&	0.26	&$	0.48\pm0.16	$&$	0.52\pm0.18	$&	-&$	0.51\pm0.20	$ &$	0.37\pm0.18	$&$	0.42\pm0.17	$&	c	\\
            &$ 0.02\pm0.12	$&$	0.08\pm0.10	$&	-	&$	-0.02\pm0.10	$&$	0.08\pm0.11	$&	-	&	-	&	-	&$	0.40\pm0.09	$&	-	&	-	&	-	&	d	\\
&	-	&	-	&	-	&	-	&	-	&	0.44	&$	0.42\pm0.16	$&$	0.39\pm0.20	$&$	1.02\pm0.25	$&$	0.40\pm0.08	$&$	0.21\pm0.12	$&	-	&	b	\\

&	0.04	&$	-0.01\pm0.11	$&$	0.02\pm0.06	$&$	-0.02\pm0.07	$&	0.09	&	0.25	&$	0.43\pm0.08	$&	0.47	&	0.52	&$	0.33\pm0.07	$&$	0.41\pm0.03	$&$	0.43\pm0.10	$&	h	\\

																																																			\hline

HR~5692 &	-	&	$-0.08\pm0.09$	&	-	&	-	&	-	&	0.32	&$	0.50\pm0.20	$&$	0.64\pm0.19	$&$	0.81\pm0.07	$&$	0.44\pm0.10	$&$	0.15\pm0.13	$&	-	&	b	\\
&$	0.08\pm0.04	$&$	0.02\pm0.11	$&$	0.04\pm0.06	$&$	-0.01\pm0.07	$&	0.17	&	0.72	&$	0.56\pm0.05	$&	-	&	0.61	&$	0.37\pm0.10	$&$	0.56\pm0.03	$&$	0.37\pm0.03	$&	h	\\

\hline
HR~1608&	-	&$	0.11\pm0.13	$&	-	&	-	&	-	&$	0.09\pm0.04	$&$	-0.12\pm0.11	$&$	0.2\pm0.18	$&$	0.18\pm0.04	$&$	-0.05\pm0.14	$&$	-0.22\pm0.08	$&$	-0.04\pm0.12	$&	e	\\
      &	-	&	-	&	-	&	-	&	-	&$	0.10\pm0.01	$&$	0.03\pm0.13	$&$	0.08\pm0.17	$&$	0.65\pm0.22	$&$	0.20\pm0.06	$&$	0.03\pm0.06	$&	-	&	b	\\

      &$	0.05\pm0.04	$&$	-0.09\pm0.10	$&$	0.03\pm0.10	$&$	-0.04\pm0.07	$&$	0.02\pm0.07	$&	0.02	&$	-0.08\pm0.01	$&	0.01	&$	0.17\pm0.10	$&	-0.08	&$	0.07\pm0.10	$&$	-0.02\pm0.07	$&	h	\\

\hline
HD~13611 &	-0.57	&	0.27	&	-0.02	&	0.13	&	0.06	&	0.11	&	-	&	0.02	&	-0.06	&	-0.05	&	-0.08	&	-0.01	&	f	\\

&$	0.05\pm0.04	$&$	-0.01\pm0.11	$&$	-0.02\pm0.06	$&$	-0.06\pm0.09	$&	0.02	&	0.34	&$	0.07\pm0.05	$&	-0.09	&	0.52	&	0.13	&$	0.46\pm0.02	$&$	0.28\pm0.03	$&	h	\\

\hline

HD~26965 &$	-0.07\pm0.03	$&	-0.31	&$	0.19\pm0.05	$&$	0.04\pm0.03	$&	-	&	-	&	-	&	-	&	-	&	-	&	-	&	-	&	g	\\

&$	-0.15\pm0.04	$&$	-0.38\pm0.11	$&$	0.21\pm0.08	$&$	0.02\pm0.06	$&$	0.07\pm0.06	$&	-0.06	&	-0.20	&	-	&$	-0.20\pm0.02	$&	-	&	-	&	-	&	h	\\

\hline
%\multicolumn{26}{l}{$a$ \citealt{2010Bubar}; $b$ \citealt{Merle2016}; $c$ \citealt{Yushchenko2004}; $d$ \citealt{Silva2015}; $e$ \citealt{Merle2014}; $f$ \citealt{Luck2014};
% $g$ \citealt{2012Adibekyan}; $h$ this work };\\
\multicolumn{14}{l}{reference: the same as Table \ref{ComStarPara}}\\
\end{tabular}
%}
\end{table}
\end{landscape}

\begin{landscape}
\begin{table}
\centering
\caption{The stellar abundances of program stars. The total uncertainties are given here. }
\label{abundance}
\scalebox{0.7}{
\begin{tabular}{l|rlrlrlrlrlrlrlrlrl}
\hline\hline
El  &
HD~13611     &
&
HD~202109  &
&
HD~218356 &
&
HR~5692 &
&
$BD-01^\circ469$  &
&
HR~1608 &
&
BD00$^\circ$4234 &
&
HD~26965  &
&
BD$-$01$^\circ$343  &

\\
%s10 \\
%s11 &
%s12 &
%s13 &
%s14  &
%s15 &
%s16 &
%s17  &
%s18  \\

 \hline
    [C/Fe]	&		-			&		&	$	-0.04 	\pm	0.09 	$&	(2)  	&		-			&		&		-			&		&	$	0.03 	\pm	0.10 	$&	(1)	&	$	-0.15 	\pm	0.08 	$&	(1)	&		-			&		&	-   &		&	$	-0.11 	\pm	0.07 	$&	(2)	\\

  [Na/Fe]	&	$	0.22 	\pm	0.08 	$&	(2)  	&	$	0.20 	\pm	0.09 	$&	(2)  	&	$	0.33 	\pm	0.10 	$&	(2)	&	$	0.04 	\pm	0.08 	$&	(2)	&	$	0.08 	\pm	0.09 	$&	(2)	&	$	-0.02 	\pm	0.09 	$&	(3)	&	$	-0.08 	\pm	0.07 	$&	(2)	&	$	0.10 	\pm	0.08 	$&	(3)	&	$	0.11 	\pm	0.07 	$&	(2)	\\

 [ Mg/Fe]	&	$	0.11 	\pm	0.08 	$&	(6)  	&	$	0.11 	\pm	0.08 	$&	(6)  	&	$	0.24 	\pm	0.07 	$&	(3)	&	$	0.05 	\pm	0.08 	$&	(4)	&	$	0.13 	\pm	0.08 	$&	(4)	&	$	0.02 	\pm	0.07 	$&	(5)	&	$	0.02 	\pm	0.07 	$&	(1)	&	$	0.27 	\pm	0.05 	$&	(3)	&	$	0.00 	\pm	0.07 	$&	(7)	\\

  [Al/Fe]	&	$	0.04 	\pm	0.09 	$&	(5)  	&	$	0.11 	\pm	0.08 	$&	(5)  	&	$	0.34 	\pm	0.08 	$&	(4)	&	$	0.01 	\pm	0.08 	$&	(3)	&	$	0.19 	\pm	0.08 	$&	(4)	&	$	0.13 	\pm	0.07 	$&	(6)	&		-			&		&	$	0.28 	\pm	0.05 	$&	(5)	&	$	0.06 	\pm	0.07 	$&	(4)	\\

  [Si/Fe]	&	$	0.00 	\pm	0.08 	$&	(20) 	&	$	0.10 	\pm	0.08 	$&	(21) 	&	$	0.06 	\pm	0.08 	$&	(14)	&	$	0.02 	\pm	0.08 	$&	(17)	&	$	0.09 	\pm	0.09 	$&	(17)	&	$	-0.05 	\pm	0.07 	$&	(23)	&	$	0.36 	\pm	0.07 	$&	(7)	&	$	0.15 	\pm	0.05 	$&	(21)	&	$	0.00 	\pm	0.06 	$&	(18)	\\

  [S/Fe]	&	$	-0.05 	\pm	0.09 	$&	(1)  	&	$	0.27 	\pm	0.10 	$&	(1)  	&		-			&		&		-			&		&		-			&		&	$	-0.09 	\pm	0.08 	$&	(3)	&		-			&		&	$	0.27 	\pm	0.07 	$&	(1)	&	$	0.04 	\pm	0.07 	$&	(1)	\\

  [K/Fe]	&	$	0.31 	\pm	0.18 	$&	(1)  	&	$	0.05 	\pm	0.28 	$&	(1)  	&		-			&		&	$	0.10 	\pm	0.25 	$&	(1)	&	$	-0.51 	\pm	0.44 	$&	(1)	&	$	0.22 	\pm	0.11 	$&	(1)	&		-			&		&	$	-0.04 	\pm	0.09 	$&	(1)	&	$	-0.34 	\pm	0.10 	$&	(1)	\\

 [ Ca/Fe]	&	$	0.08 	\pm	0.10 	$&	(6)  	&	$	0.05 	\pm	0.10 	$&	(7)  	&	$	0.17 	\pm	0.10 	$&	(3)	&	$	0.03 	\pm	0.10 	$&	(6)	&	$	-0.02 	\pm	0.10 	$&	(8)	&	$	0.13 	\pm	0.09 	$&	(16)	&	$	0.12 	\pm	0.10 	$&	(3)	&	$	0.16 	\pm	0.07 	$&	(5)	&	$	-0.11 	\pm	0.08 	$&	(5)	\\

[Sc/Fe]	&	$	-0.05 	\pm	0.11 	$&	(4)  	&	$	-0.01 	\pm	0.11 	$&	(8)  	&	$	0.04 	\pm	0.12 	$&	(3)	&	$	0.04 	\pm	0.10 	$&	(5)	&	$	0.08 	\pm	0.11 	$&	(6)	&	$	-0.10 	\pm	0.09 	$&	(8)	&	$	0.08 	\pm	0.08 	$&	(3)	&	$	0.19 	\pm	0.08 	$&	(6)	&	$	0.04 	\pm	0.10 	$&	(9)	\\

[Ti/Fe]	&	$	0.02 	\pm	0.14 	$&	(15) 	&	$	-0.04 	\pm	0.12 	$&	(31) 	&	$	0.01 	\pm	0.13 	$&	(5)	&	$	0.06 	\pm	0.12 	$&	(14)	&	$	-0.07 	\pm	0.15 	$&	(10)	&	$	0.03 	\pm	0.10 	$&	(29)	&	$	0.07 	\pm	0.10 	$&	(11)	&	$	0.24 	\pm	0.09 	$&	(23)	&	$	-0.15 	\pm	0.10 	$&	(27)	\\

  [ V/Fe]	&	$	-0.03 	\pm	0.11 	$&	(5)  	&	$	0.06 	\pm	0.12 	$&	(7)  	&-	&		&	$	0.14 	\pm	0.12 	$&	(7)	&	$	0.26 	\pm	0.16 	$&	(5)	&	$	0.09 	\pm	0.09 	$&	(7)	&	$	0.06 	\pm	0.10 	$&	(8)	&	$	0.43 	\pm	0.09 	$&	(10)	&	$	0.25 	\pm	0.12 	$&	(3)	\\

[Cr/Fe]	&	$	-0.01 	\pm	0.11 	$&	(5)  	&	$	0.05 	\pm	0.11 	$&	(15) 	&	$	0.07 	\pm	0.11 	$&	(5)	&	$	0.03 	\pm	0.11 	$&	(11)	&	$	-0.02 	\pm	0.11 	$&	(9)	&	$	0.00 	\pm	0.09 	$&	(14)	&	$	-0.04 	\pm	0.08 	$&	(5)	&	$	0.06 	\pm	0.07 	$&	(5)	&	$	-0.07 	\pm	0.08 	$&	(14)	\\

 [ Mn/Fe]	&	$	0.05 	\pm	0.14 	$&	(3)  	&	$	0.04 	\pm	0.09 	$&	(2)  	&	$	0.09 	\pm	0.11 	$&	(3)	&	$	0.08 	\pm	0.11 	$&	(3)	&	$	0.30 	\pm	0.10 	$&	(1)	&	$	0.05 	\pm	0.11 	$&	(3)	&	$	-0.41 	\pm	0.08 	$&	(3)	&	$	-0.15 	\pm	0.08 	$&	(4)	&	$	0.13 	\pm	0.07 	$&	(2)	\\

[Fe/H]	&	$	-0.01 	\pm	0.07 	$&	(127)	&	$	-0.01 	\pm	0.10 	$&	(155)	&	$	-0.43 	\pm	0.06 	$&	(80)	&	$	0.02 	\pm	0.07 	$&	(136)	&	$	-0.09 	\pm	0.07 	$&	(135)	&	$	-0.09 	\pm	0.09 	$&	(191)	&	$	-0.96 	\pm	0.05 	$&	(82)	&	$	-0.38 	\pm	0.06 	$&	(153)	&	$	0.36 	\pm	0.05 	$&	(117)	\\

  [Co/Fe]	&	$	-0.02 	\pm	0.10 	$&	(7)  	&	$	0.02 	\pm	0.10 	$&	(8)  	&	$	0.05 	\pm	0.09 	$&	(5)	&	$	0.04 	\pm	0.09 	$&	(7)	&	$	-0.01 	\pm	0.10 	$&	(2)	&	$	0.03 	\pm	0.09 	$&	(8)	&	$	-0.06 	\pm	0.07 	$&	(1)	&	$	0.21 	\pm	0.07 	$&	(10)	&	$	0.09 	\pm	0.07 	$&	(6)	\\

[Ni/Fe]	&	$	-0.06 	\pm	0.09 	$&	(33) 	&	$	-0.02 	\pm	0.09 	$&	(35) 	&	$	-0.05 	\pm	0.08 	$&	(16)	&	$	-0.01 	\pm	0.09 	$&	(38)	&	$	0.08 	\pm	0.09 	$&	(31)	&	$	-0.04 	\pm	0.08 	$&	(42)	&	$	-0.26 	\pm	0.09 	$&	(4)	&	$	0.02 	\pm	0.06 	$&	(38)	&	$	-0.03 	\pm	0.07 	$&	(31)	\\

 [Cu/Fe]	&	$	0.02 	\pm	0.10 	$&	(1)  	&	$	0.09 	\pm	0.10 	$&	(1)  	&	$	0.39 	\pm	0.11 	$&	(1)	&	$	0.17 	\pm	0.11 	$&	(1)	&	$	0.17 	\pm	0.11 	$&	(1)	&	$	0.02 	\pm	0.11 	$&	(4)	&		-			&		&	$	0.07 	\pm	0.07 	$&	(3)	&	$	0.06 	\pm	0.09 	$&	(5)	\\

 [Sr/Fe]	&	$	0.34 	\pm	0.21 	$&	(1)  	&	$	0.25 	\pm	0.31 	$&	(1)  	&	$	1.09 	\pm	0.09 	$&	(1)	&	$	0.72 	\pm	0.26 	$&	(1)	&		-			&		&	$	0.02 	\pm	0.12 	$&	(1)	&		-			&		&	$	-0.06 	\pm	0.10 	$&	(1)	&	$	-0.13 	\pm	0.13 	$&	(1)	\\

   [Y/Fe]	&	$	0.07 	\pm	0.11 	$&	(5)  	&	$	0.43 	\pm	0.12 	$&	(7)  	&	$	0.74 	\pm	0.10 	$&	(3)	&	$	0.56 	\pm	0.11 	$&	(9)	&	$	0.07 	\pm	0.09 	$&	(1)	&	$	-0.08 	\pm	0.11 	$&	(2)	&	$	0.11 	\pm	0.08 	$&	(1)	&	$	-0.20 	\pm	0.07 	$&	(1)	&	$	-0.22 	\pm	0.09 	$&	(1)	\\

[Zr/Fe]	&	$	-0.09 	\pm	0.13 	$&	(1)  	&	$	0.47 	\pm	0.14 	$&	(4)  	&	-&		&		-			&		&		-			&		&	$	0.01 	\pm	0.08 	$&	(1)	&		-			&		&		-			&		&		-			&		\\

 [Ba/Fe]	&	$	0.52 	\pm	0.15 	$&	(1)  	&	$	0.52 	\pm	0.14 	$&	(1)  	&	$	1.01 	\pm	0.11 	$&	(1)	&	$	0.61 	\pm	0.12 	$&	(1)	&	$	-0.07 	\pm	0.15 	$&	(1)	&	$	0.17 	\pm	0.13 	$&	(3)	&	$	-0.25 	\pm	0.09 	$&	(1)	&	$	-0.20 	\pm	0.07 	$&	(3)	&	$	-0.15 	\pm	0.11 	$&	(2)	\\

  [La/Fe]	&	$	0.13 	\pm	0.09 	$&	(1)  	&	$	0.33 	\pm	0.10 	$&	(5)  	&	$	0.53 	\pm	0.10 	$&	(3)	&	$	0.37 	\pm	0.10 	$&	(7)	&		-			&	(1)	&	$	-0.08 	\pm	0.09 	$&	(3)	&		-			&		&		-			&		&		-			&		\\

 [Ce/Fe]	&	$	0.46 	\pm	0.16 	$&	(2)  	&	$	0.41 	\pm	0.11 	$&	(4)  	&	$	0.52 	\pm	0.12 	$&	(2)	&	$	0.56 	\pm	0.16 	$&	(2)	&	$	-0.05 	\pm	0.12 	$&	(2)	&	$	0.07 	\pm	0.09 	$&	(2)	&	$	0.15 	\pm	0.08 	$&	(1)	&		-			&		&	$	0.13 	\pm	0.09 	$&	(2)	\\

 [ Nd/Fe]	&	$	0.28 	\pm	0.10 	$&	(2)  	&	$	0.43 	\pm	0.11 	$&	(5) 	&	$	0.70 	\pm	0.11 	$&	(2)	&	$	0.37 	\pm	0.10 	$&	(3)	&	$	-0.01 	\pm	0.10 	$&	(1)	&	$	-0.02 	\pm	0.09 	$&	(3)	&		-			&		&		-			&		&	$	-0.11 	\pm	0.08 	$&	(2)	\\

\hline
%\multicolumn{10}{l}{The stars codes in the header are: s1 - HD~13611; s2 - HD~202109; s3 - HD~218356; s4 - HR~5692; s5 - BD$-$01\,469; s6 - HR~1608; s7 - BD$-$00\,4234; }\\
%%\multicolumn{10}{l}{  s10 - BD$-$01\,407; s11 - BD+39\,539; s12 - BD$-$07\,5906;s13 - BD+33\,2834 ; s14 - BD+13\,99; s15 - BD+71\,380; s16 - BD+20\,5125; s17 - BD+30\,2592;}\\
%\multicolumn{10}{l}{s8 - HD~26965; s9 - BD$-$01\,343;}
\end{tabular}
}
\end{table}
\end{landscape}

\begin{landscape}
\begin{table}
\centering
\contcaption{The differential abundances of program stars. The total uncertainties are given here. }
\label{abundance}
\scalebox{0.7}{
\begin{tabular}{l|rlrlrlrlrlrlrlrlrl}
\hline\hline
el     &
%s1  &
%s2 &
%s3 &
%s4  &
%s5 &
%s6 &
%s7  &
%s8  &
%s9 &
 BD$-$01$^\circ$407 &
    &
BD+39$^\circ$539 &
    &
BD$-$07$^\circ$5906 &
    &
BD+33$^\circ$2834 &
    &
BD+13$^\circ$99  &
    &
BD+71$^\circ$380 &
    &
BD+20$^\circ$5125 &
     &
BD+30$^\circ$2592 &
     &
HD~39570 &
  \\

 \hline
        [C/Fe]	&	$	0.15 	\pm	0.06 	$	&	(1)	&	-	&	 	&	$	0.19 	\pm	0.08 	$	&	(3)   	&	$	-0.15 	\pm	0.07 	$&	(4)  	&		-			&		&	$	0.09 	\pm	0.07 	$	&	(1)  	&	-     &	   	&		-			&		&	$	-0.06 	\pm	0.07 	$&	(4)  	\\

  [Na/Fe]	&	$	-0.12 	\pm	0.05 	$	&	(2)	&	$	-0.01 	\pm	0.08 	$	&	(1) 	&	$	-0.06 	\pm	0.07 	$	&	(1)   	&	$	-0.09 	\pm	0.08 	$&	(3)  	&	$	-0.08 	\pm	0.06 	$&	(3)  	&	$	0.04 	\pm	0.07 	$	&	(3)  	&	$	0.04 	\pm	0.09 	$&	(2)    	&	$	0.03 	\pm	0.06 	$&	(2)  	&	$	-0.05 	\pm	0.08 	$&	(3)  	\\

 [ Mg/Fe]	&	$	-0.09 	\pm	0.05 	$	&	(4)	&	$	-0.07 	\pm	0.08 	$	&	(3) 	&	$	-0.05 	\pm	0.09 	$	&	(4)   	&	$	-0.04 	\pm	0.07 	$&	(4)  	&	$	-0.05 	\pm	0.05 	$&	(6)  	&	$	0.20 	\pm	0.06 	$	&	(6)  	&	$	-0.05 	\pm	0.07 	$&	(4)    	&	$	-0.06 	\pm	0.07 	$&	(5)  	&	$	0.02 	\pm	0.08 	$&	(6)  	\\

  [Al/Fe]	&	$	-0.04 	\pm	0.06 	$	&	(4)	&	$	-0.05 	\pm	0.06 	$	&	(6) 	&	$	0.09 	\pm	0.08 	$	&	(3)   	&	$	-0.06 	\pm	0.06 	$&	(4)  	&	$	0.04 	\pm	0.05 	$&	(5)  	&	$	0.18 	\pm	0.06 	$	&	(4)  	&	$	0.01 	\pm	0.07 	$&	(3)    	&	$	0.08 	\pm	0.06 	$&	(5)  	&	$	0.13 	\pm	0.07 	$&	(6)  	\\

  [Si/Fe]	&	$	-0.03 	\pm	0.05 	$	&	(23)	&	$	0.01 	\pm	0.07 	$	&	(11)	&	$	0.09 	\pm	0.07 	$	&	(18)  	&	$	0.02 	\pm	0.06 	$&	(20) 	&	$	-0.03 	\pm	0.05 	$&	(21) 	&	$	0.14 	\pm	0.05 	$	&	(21) 	&	$	0.04 	\pm	0.06 	$&	(19)   	&	$	-0.02 	\pm	0.06 	$&	(20) 	&	$	0.05 	\pm	0.07 	$&	(25) 	\\

  [S/Fe]	&	$	0.09 	\pm	0.06 	$	&	(1)	&		-				&	    	&	$	0.22 	\pm	0.08 	$	&	(3)   	&	$	-0.07 	\pm	0.07 	$&	(2)  	&		-			&	     	&		-				&	     	&		-			&	       	&		-			&	     	&	$	0.00 	\pm	0.08 	$&	(3)  	\\

  [K/Fe]	&	$	-0.18 	\pm	0.09 	$	&	(1)	&	$	-0.44 	\pm	0.10 	$	&	(1) 	&	$	0.02 	\pm	0.10 	$	&	(1)   	&	$	0.23 	\pm	0.09 	$&	(1)  	&	$	-0.09 	\pm	0.09 	$&	(1)  	&	$	0.32 	\pm	0.09 	$	&	(1)  	&	$	-0.42 	\pm	0.12 	$&	(1)    	&	$	-0.13 	\pm	0.10 	$&	(1)  	&	$	0.18 	\pm	0.09 	$&	(1)  	\\

 [ Ca/Fe]	&	$	0.01 	\pm	0.08 	$	&	(5)	&	$	-0.07 	\pm	0.09 	$	&	(2) 	&	$	0.11 	\pm	0.08 	$	&	(9)   	&	$	0.09 	\pm	0.07 	$&	(14) 	&	$	-0.02 	\pm	0.07 	$&	(10) 	&	$	0.23 	\pm	0.08 	$	&	(11) 	&	$	-0.12 	\pm	0.07 	$&	(4)    	&	$	0.03 	\pm	0.07 	$&	(5)  	&	$	0.08 	\pm	0.08 	$&	(18) 	\\

[Sc/Fe]	&	$	-0.04 	\pm	0.07 	$	&	(4)	&	$	0.02 	\pm	0.09 	$	&	(3) 	&	$	0.13 	\pm	0.08 	$	&	(3)   	&	$	-0.10 	\pm	0.08 	$&	(4)  	&	$	-0.01 	\pm	0.08 	$&	(7)  	&	$	0.05 	\pm	0.08 	$	&	(4)  	&	$	0.04 	\pm	0.08 	$&	(7)    	&	$	-0.05 	\pm	0.08 	$&	(6)  	&	$	0.04 	\pm	0.09 	$&	(5)  	\\

[Ti/Fe]	&	$	-0.02 	\pm	0.09 	$	&	(31)	&	$	-0.17 	\pm	0.11 	$	&	(9) 	&	$	0.02 	\pm	0.09 	$	&	(13)  	&	$	0.01 	\pm	0.07 	$&	(22) 	&	$	0.03 	\pm	0.08 	$&	(27) 	&	$	0.24 	\pm	0.08 	$	&	(27) 	&	$	-0.05 	\pm	0.10 	$&	(21)   	&	$	-0.02 	\pm	0.09 	$&	(27) 	&	$	0.02 	\pm	0.08 	$&	(31) 	\\

  [ V/Fe]	&	$	0.08 	\pm	0.09 	$	&	(8)	&	$	0.18 	\pm	0.11 	$	&	(4) 	&	$	-0.04 	\pm	0.09 	$	&	(6)   	&	$	0.02 	\pm	0.07 	$&	(3)  	&	$	0.09 	\pm	0.08 	$&	(6)  	&	$	0.16 	\pm	0.09 	$	&	(5)  	&	$	0.33 	\pm	0.11 	$&	(7)    	&	$	0.18 	\pm	0.09 	$&	(7)  	&	$	-0.02 	\pm	0.08 	$&	(6)  	\\

[Cr/Fe]	&	$	0.00 	\pm	0.07 	$	&	(13)	&	$	-0.13 	\pm	0.09 	$	&	(7) 	&	$	0.04 	\pm	0.09 	$	&	(10)  	&	$	0.02 	\pm	0.07 	$&	(16) 	&	$	-0.03 	\pm	0.07 	$&	(10) 	&	$	0.04 	\pm	0.07 	$	&	(14) 	&	$	-0.12 	\pm	0.07 	$&	(8)    	&	$	-0.01 	\pm	0.07 	$&	(9)  	&	$	0.03 	\pm	0.08 	$&	(15) 	\\

 [ Mn/Fe]	&	$	-0.10 	\pm	0.08 	$	&	(4)	&	$	-0.07 	\pm	0.10 	$	&	(4) 	&	$	-0.10 	\pm	0.10 	$	&	(4)   	&	$	-0.20 	\pm	0.08 	$&	(15) 	&	$	-0.15 	\pm	0.10 	$&	(3)  	&	$	-0.25 	\pm	0.08 	$	&	(4)  	&	$	-0.14 	\pm	0.06 	$&	(2)    	&	$	-0.02 	\pm	0.05 	$&	(4)  	&	$	-0.10 	\pm	0.09 	$&	(5)  	\\

[Fe/H]	&	$	-0.12 	\pm	0.04 	$	&	(154)	&	$	0.01 	\pm	0.07 	$	&	(54)	&	$	-0.13 	\pm	0.06 	$	&	(131) 	&	$	-0.06 	\pm	0.07 	$&	(168)	&	$	-0.38 	\pm	0.06 	$&	(164)	&	$	-0.41 	\pm	0.05 	$	&	(140)	&	$	-0.14 	\pm	0.06 	$&	(127)  	&	$	-0.07 	\pm	0.07 	$&	(153)	&	$	0.03 	\pm	0.08 	$&	(205)	\\

  [Co/Fe]	&	$	-0.04 	\pm	0.06 	$	&	(8)	&	$	0.16 	\pm	0.07 	$	&	(8) 	&	$	-0.07 	\pm	0.08 	$	&	(5)   	&	$	-0.07 	\pm	0.06 	$&	(2)  	&	$	-0.02 	\pm	0.06 	$&	(7)  	&	$	0.09 	\pm	0.07 	$	&	(5)  	&	$	0.08 	\pm	0.06 	$&	(9)    	&	$	0.05 	\pm	0.09 	$&	(5)  	&	$	-0.02 	\pm	0.07 	$&	(5)  	\\

[Ni/Fe]	&	$	-0.10 	\pm	0.06 	$	&	(30)	&	$	-0.01 	\pm	0.07 	$	&	(27)	&	$	-0.04 	\pm	0.07 	$	&	(36)  	&	$	-0.04 	\pm	0.07 	$&	(34) 	&	$	-0.07 	\pm	0.06 	$&	(35) 	&	$	0.02 	\pm	0.07 	$	&	(34) 	&	$	-0.06 	\pm	0.06 	$&	(37)   	&	$	-0.05 	\pm	0.07 	$&	(42) 	&	$	-0.01 	\pm	0.07 	$&	(40) 	\\

 [Cu/Fe]	&	$	-0.08 	\pm	0.08 	$	&	(3)	&		-				&		&	$	-0.05 	\pm	0.10 	$	&	(4)   	&	$	-0.35 	\pm	0.07 	$&	(3)  	&	$	-0.10 	\pm	0.06 	$&	(2)  	&	$	-0.07 	\pm	0.08 	$	&	(2)  	&	$	-0.13 	\pm	0.08 	$&	(2)    	&	$	-0.05 	\pm	0.08 	$&	(3)  	&	$	-0.06 	\pm	0.10 	$&	(3)  	\\

 [Sr/Fe]	&	$	-0.03 	\pm	0.11 	$	&	(1)	&	$	-0.19 	\pm	0.12 	$	&	(1) 	&	$	-0.20 	\pm	0.09 	$	&	(1)   	&	$	0.12 	\pm	0.07 	$&	(1)  	&		-			&	     	&	$	-0.04 	\pm	0.08 	$	&	(1)  	&	$	0.00 	\pm	0.12 	$&	(1)    	&	$	0.14 	\pm	0.11 	$&	(1)  	&	$	-0.11 	\pm	0.08 	$&	(1)  	\\

   [Y/Fe]	&	$	0.09 	\pm	0.08 	$	&	(2)	&		-				&	    	&	$	-0.15 	\pm	0.09 	$	&	(3)   	&	$	0.05 	\pm	0.08 	$&	(4)  	&	$	-0.09 	\pm	0.07 	$&	(1)  	&	$	-0.16 	\pm	0.08 	$	&	(3)  	&	$	-0.31 	\pm	0.07 	$&	(1)    	&	$	-0.06 	\pm	0.08 	$&	(1)  	&	$	-0.12 	\pm	0.09 	$&	(5)  	\\

[Zr/Fe]	&		-				&		&		-				&		&	$	0.08 	\pm	0.08 	$	&	(1)   	&	$	-0.32 	\pm	0.07 	$&	(1)  	&	$	-0.16 	\pm	0.07 	$&	(1)  	&		-				&	     	&-	&	   	&		-			&	     	&	$	-0.29 	\pm	0.08 	$&	(1)  	\\

 [Ba/Fe]	&	$	0.07 	\pm	0.10 	$	&	(2)	&	$	-0.18 	\pm	0.10 	$	&	(3) 	&	$	-0.01 	\pm	0.10 	$	&	(2)   	&	$	0.17 	\pm	0.09 	$&	(4)  	&	$	-0.04 	\pm	0.08 	$&	(3)  	&	$	-0.09 	\pm	0.09 	$	&	(3)  	&	$	-0.19 	\pm	0.08 	$&	(2)    	&	$	-0.12 	\pm	0.09 	$&	(2)  	&	$	-0.05 	\pm	0.10 	$&	(3)  	\\

  [La/Fe]	&		-				&		&		-				&	    	&		-				&	      	&		-			&	     	&		-			&	     	&	$	0.15 	\pm	0.07 	$	&	(1)  	&-&	   	&	$	-0.08 	\pm	0.08 	$&	(1)  	&	-  &	  	\\

 [Ce/Fe]	&	$	0.24 	\pm	0.07 	$	&	(1)	&	$	0.10 	\pm	0.10 	$	&	(2) 	&	$	0.01 	\pm	0.09 	$	&	(2)   	&	$	0.12 	\pm	0.08 	$&	(2)  	&	$	0.14 	\pm	0.07 	$&	(1)  	&	$	0.02 	\pm	0.08 	$	&	(1)  	&	- &	   	&	$	0.16 	\pm	0.08 	$&	(2)  	&	$	0.01 	\pm	0.09 	$&	(2)  	\\

 [ Nd/Fe]	&	$	0.16 	\pm	0.08 	$	&	(2)	&		-				&		&	$	-0.11 	\pm	0.09 	$	&	(1)   	&		-			&		&	$	0.17 	\pm	0.07 	$&	(1)  	&		-				&		&	$	0.95 	\pm	0.07 	$&	(1)    	&	$	-0.10 	\pm	0.09 	$&	(1)  	&	$	-0.13 	\pm	0.08 	$&	(1)  	\\

\hline
%\multicolumn{10}{l}{The stars codes in the header are: s1 - HD~13611; s2 - HD~202109; s3 - HD~218356; s4 - HR~5692; s5 - BD$-$01\,469; s6 - HR~1608; s7 - BD$-$00\,4234; s8 - HD~26965;}\\
%s9 - BD$-$01\,343;
%\multicolumn{10}{l}{  s10 - BD$-$01\,407; s11 - BD+39\,539; s12 - BD$-$07\,5906;s13 - BD+33\,2834 ; s14 - BD+13\,99; s15 - BD+71\,380; s16 - BD+20\,5125; s17 - BD+30\,2592;}\\
%\multicolumn{10}{l}{  s18 - HD~39570.}
\end{tabular}
}
\end{table}
\end{landscape}

\section*{Acknowledgements}
 We are thankful to the referee for their constructive comments and suggestions that led to us improving the manuscript. We thank Ji Li for providing help for the Observation. We also thank Yujuan Liu for valuable comments and many stimulating discussions. This study was supported by the National Natural Science Foundation of China under grant number 11390371, 11233004 and U1431106. YBK thanks the Chinese Academy of Sciences (CAS) for support through a CAS PIFI fellowship.

%\bibliographystyle{mnras}
%%\begin{thebibliography}{}
\bibliography{bibliography}

\section{Some extra material}

%%%%%%%%%%%%%%%%%%%%%%%%%%%%%%%%%%%%%%%%%%%%%%%%%%

% Don't change these lines
\bsp	% typesetting comment
\label{lastpage}

\end{document}